\documentclass[twocolumn,floatfix,10pt,superscriptaddress,prx,longbibliography]{revtex4-2}
\usepackage{graphicx}
\usepackage{dcolumn} 
\usepackage[utf8]{inputenc} 
\usepackage[T1]{fontenc}
\usepackage{newtx}
\usepackage{etoolbox}
\usepackage{natbib}
\usepackage{latexsym}
\usepackage{amsmath}
\usepackage{amsfonts}
\usepackage{gensymb}
\usepackage{hyperref}
\usepackage{booktabs}
\usepackage{braket}
\usepackage{bbold}
\usepackage{slashed}
\usepackage[dvipsnames]{xcolor}
\usepackage[normalem]{ulem}
\usepackage{siunitx}
\usepackage[version=4]{mhchem}
\preprint{APS/123-QED}

\newcommand{\st}[1]{\left\{#1\right\}}

\renewcommand{\v}[1]{\boldsymbol{#1}}

\newcommand{\supplementarysection}{%
  \setcounter{figure}{0}
  \let\oldthefigure\thefigure
  \renewcommand{\thefigure}{S\oldthefigure}
  \setcounter{section}{0}
  \let\oldthesection\thesection
  \renewcommand{\thesection}{S\oldthesection}
  \setcounter{equation}{0}
  \let\oldtheequation\theequation
  \renewcommand{\theequation}{S\oldtheequation}
}

\begin{document}

\title{Dynamic twisting and imaging of moiré crystals}

\author{Qixuan Zhang}
\affiliation{Program in Material Science and Engineering, University of California, San Diego, CA 92093, USA.}
\author{Lingyuan Lyu}
\author{Sneh Pancholi}
\author{Ziying Yan}
\author{Trevor Senaha}
\author{Ruolun Zhang}
\author{Chen Wu}
\author{Leonard W. Cao}
\affiliation{Department of Physics, University of California, San Diego, CA 92093, USA.}

\author{Jason Tresback}
\affiliation{Center for Nanoscale Systems, Harvard University, Cambridge, MA 02138, USA.}

\author{Andrew Dai}
\affiliation{Department of Physics, University of California, San Diego, CA 92093, USA.}

\author{Kenji Watanabe}
\affiliation{Research Center for Electronic and Optical Materials, National Institute for Material Science, Namiki 1-1, Tsukuba, Ibaraki 305-0044, Japan.}
\author{Takashi Taniguchi}
\affiliation{Research Center for Materials Nanoarchitectonics, National Institute for Material Science, Namiki 1-1, Tsukuba, Ibaraki 305-0044, Japan.}

\author{Daniel E. Parker}
\affiliation{Department of Physics, University of California, San Diego, CA 92093, USA.}

\author{Monica T. Allen}
\affiliation{Department of Physics, University of California, San Diego, CA 92093, USA.}


\begin{abstract}
Moiré superlattices in stacked 2D crystals are powerful platforms for engineering correlated and topological quantum phases, with twisted graphene and transition metal dichalcogenides (TMDs) as prominent examples. Their angle-sensitive band structures enable rich tunability; however, conventional tear-and-stack methods fix the angle at assembly, limiting systematic exploration of angle-dependent phenomena. Here, we present a scanning-probe-based manipulation scheme that enables \textit{in situ}, continuous post-fabrication twist control using nanostructured metal rotors. We demonstrate reproducible angle tuning and direct moiré imaging across three platforms: graphene, hBN, and encapsulated, air-sensitive \ce{MoTe2}. Quantitative piezoresponse force microscopy (PFM) analysis confirms sub-degree precision with minimal induced heterostrain, preserving sample quality even in the marginally twisted regime. Crucially, the device architecture maintains open access to the active region, allowing optical, scanning-probe, and transport measurements. This work enables single-device mapping of the angular phase diagram of moir\'e material including the minimally twisted regime.
\end{abstract}

\maketitle


\begin{figure*}
\includegraphics[width= 0.9\textwidth]{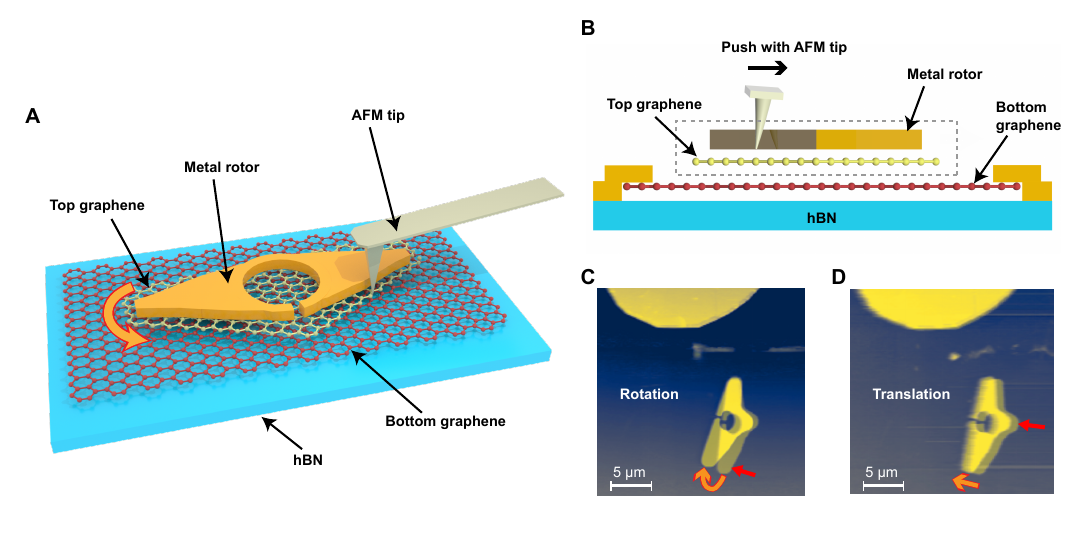}
\caption{\label{fig:Schematic} \textbf{Mechanical manipulation of a moiré superlattice, enabling \textit{in situ} control over the twist angle.}
\textbf{(A)} Schematic illustration of a twisted bilayer graphene rotor device and dynamic manipulation of the twist angle using an AFM tip. The metal rotor is clamped onto the top graphene layer, so that when the AFM tip pushes the rotor, then the top graphene moves relative to the bottom graphene. The orange arrow indicates the induced motion of the rotor under the applied AFM force. The circular opening in the middle of the rotor provides access to the moiré superlattice for, e.g., high-resolution imaging. 
\textbf{(B)} Cross-sectional view of the tip-induced mechanical manipulation of the rotor frame during a contact-mode AFM line scan.
\textbf{(C)} Tapping-mode AFM overlay of the rotor device before and after rotation. The red arrow shows the direction of lateral force from the AFM tip, and the orange arrow is the resulting clockwise rotation.
\textbf{(D)} Tapping-mode AFM overlay of the rotor before and after lateral translation of the top graphene relative to the bottom graphene. The red arrow indicates the AFM tip's lateral force direction. 
}
\end{figure*}


The electronic structure of van der Waals moir\'e heterostructures is exceptionally sensitive to the relative crystallographic alignment of their constituent atomic layers. Small interlayer rotations generate long-wavelength moir\'e potentials that reshape band dispersion and topology, enabling interaction-driven ground states across a broad range of material platforms \cite{Andrei2021NatRevMatMarvels,Carr2020NatRevMatMethods,Nuckolls2024NatRevMatMicroscopic,Yankowitz2012NatPhysSuperlatticeDirac,Dean2013NatureHofstadter,Ponomarenko2013NatureCloning,Hunt2013ScienceMassiveDirac}. 
In twisted bilayer graphene (TBG), narrow bands near the ``magic angle'' host correlated insulators, unconventional superconductivity, and quantum anomalous Hall states \cite{Andrei_2020, Nimbalkar2020OpportunitiesAC, Wangmoire2021, Nuckolls2024AMP}. Beyond graphene, moir\'e phenomena have been observed in twisted transition metal dichalcogenides (tTMDs), whose strong spin–orbit coupling, broken inversion symmetry, and valley-contrasting physics support correlated and topological phases over a wide span of twist angles
\cite{wu2019topological,pan2020quantum,zeng2023integer,cai2023signatures,mak2022semiconductor}. More broadly, moir\'e periodicity provides a versatile design principle for applications ranging from photonic superlattices to angle-tunable electrochemical platforms \cite{Du2023_MoirePhotonics_Science,Zhang2023_TwistedPhC_Vortex_NatCommun,Nguyen2022_MagicConfigs_PRResearch,Yu2022_TwistElectrochem_NatChem}. A central factor enabling such progress is the high tunability of moir\'e devices, whose displacement field, carrier density and magnetic field can be adjusted post-fabrication.

However, the structural parameters that most directly determine the underlying moir\'e Hamiltonian --- such as the global twist angle,  heterostrain, and disorder --- are typically fixed at the time of fabrication. Conventional ``tear-and-stack" fabrication permanently fixes the interlayer twist angle, forcing systematic exploration of angle space to rely on ensembles of nominally similar devices that inevitably differ in disorder, strain and dielectric environment. These constraints consequently hinder experimental reproducibility and obscure the interpretation of intrinsic angle-dependent physics \cite{dean2010boron,geim2013van,edgecontacts2013,tearstack2016}.

These practical limitations have motivated substantial experimental efforts to precisely read out and tune the geometric structure of moir\'e superlattices. 
There are two fundamental challenges that lie at the heart of any \textit{in situ} twisting technology: (1) the challenge of \textit{uniform control}: precisely rotating a single atomic layer relative to another in a uniform manner across the sample, and without introducing strain or other mesoscopic disorder, and (2) the challenge of \textit{readout}: high-resolution imaging of the superlattice structure in real space, which can directly resolve moir\'e lattice reconstruction, local twist angle variations, heterostrain fields, and domain wall networks at atomic to nanometer length scales \cite{Yoo2019NatMater,Weston2020NatNanotechTMDreconstruction,Kazmierczak2021_StrainTBG_NatMater,VanWinkle2023NatCommunReconstructionTMD}. Because these microscopic effects significantly impact band structure, real-space superlattice characterization has been crucial for the development of accurate theoretical models of emergent phenomena in moir\'e systems
\cite{Yoo2019NatMater,Weston2020NatNanotechTMDreconstruction,Kazmierczak2021_StrainTBG_NatMater,Luo2020NatCommunNanoIR,deJong2022_MoireDeformation_TBG_NatCommun, Uri2020_Nature, Kerelsky2019_Nature, Kazmierczak2021_NatMater}.

Over the past 10 years, great strides have been made on the development of rotatable van der Waals structures 
\cite{koren2016coherent,ribeiro2018twistable,Yang2020SciAdv,arrighi2023nonidentical,farrar2025impact,inbar2023quantum,kapfer2023programming,Tang2024MEGA2D,Birkbeck2025QTMPhononsTBG}, but at the cost of either uniformity or spatially-resolved readout of the superlattice.
For instance, nanomechanical bending was used to generate a twist angle gradient in TBG nanoribbons
\cite{kapfer2022programming}. While this indeed allowed mechanical reconfiguration, it was accompanied by large strains and twist angle disorder that strongly broke superlattice translation symmetry.
A contrasting approach has achieved rotation of two independently-contacted graphite or graphene devices against one another
using an AFM tip \cite{koren2016coherent} or micropositoners \cite{inbar2023quantum}. However, because the upper and lower layers are mounted onto thick substrates, the moir\'e superlattice at the interface cannot be directly imaged, hindering readout.

Given the sensitivity of  moir\'e systems to the microscopic details of the superlattice, it is of critical importance to meet \textit{both} requirements of uniformity and readout in the same device. This is the key objective of the present work.
Here we introduce a scanning-probe-based manipulation platform that enables \textit{in situ} twisting and imaging of the moir\'e superlattice in fully fabricated devices. This approach uses a lithographically patterned windowed metal rotor rigidly coupled to the top layer and actuated by programmed lateral forces from an atomic force microscope (AFM) tip, producing precise sub-degree rotation with global twist angle uniformity and minimal heterostrain. The rotor platform provides five key capabilities: (I) \textit{in situ} post-fabrication reconfigurability, enabling repeatable, stepwise tuning of the global twist angle within a single device; (II) control of a single atomic layer within an heterostructure; (III) access wide twist angle regime of moir\'e devices with low strain and angular disorder; (IV) microscopic structural verification, enabled by an open measurement window that permits high-resolution imaging and validation of the evolving superlattice geometry after each tuning step; and (V) broad material compatibility, including homobilayers, heterobilayers, and fully encapsulated homobilayers of air-sensitive tTMDs.

We demonstrate simultaneous \textit{in situ} rotation and imaging in two moir\'e systems of significant interest: twisted bilayer graphene (TBG) and twisted \ce{MoTe2} (\ce{tMoTe2}). Using piezoresponse force microscopy (PFM), we image the moir\'e lattice before and after each actuation step and quantitatively extract the local twist-angle distribution and heterostrain. These measurements confirm successive twist-angle control with minimal induced heterostrain and low angular disorder, establishing a route to systematically explore the twist angle dependence of moir\'e systems with simultaneous control and characterization of the lattice geometry and quality required for meaningful comparison between experiment and theory.

\begin{figure*}
  \centering
  \includegraphics[width=0.9\textwidth]{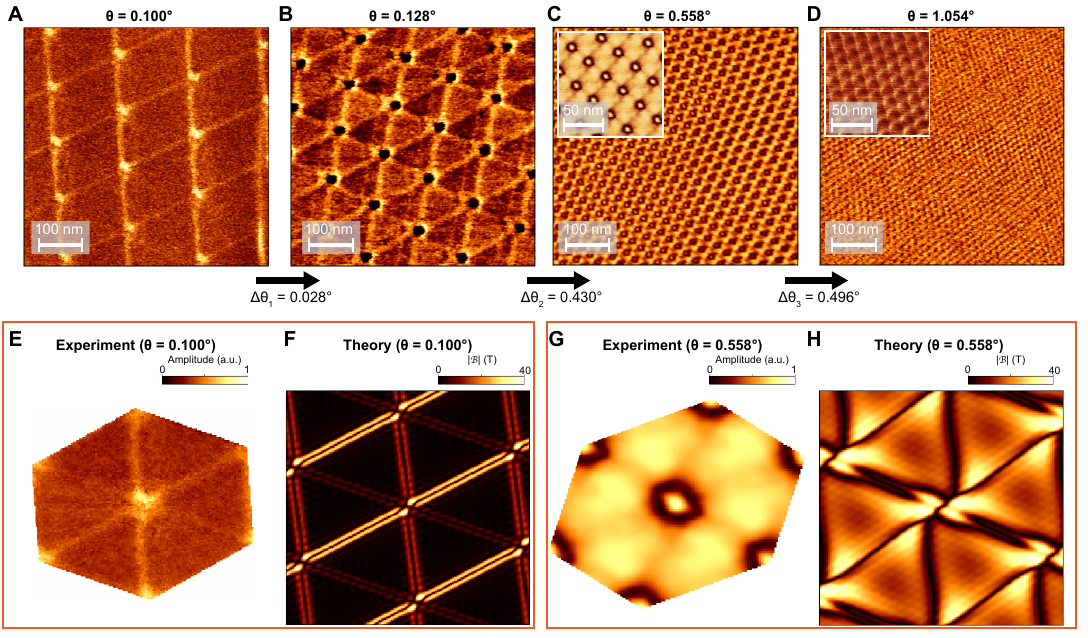}
  \caption{\label{fig:fig2}
  \textbf{
  Nanoscale tuning of the moiré superlattice via \textit{in situ} rotation of TBG.}
  \textbf{(A–D)} PFM images before and after sequential AFM-driven rotations: initial \(\theta=0.100^\circ\); first push \(\Delta\theta_1=0.028^\circ\) (\(\theta=0.128^\circ\)); subsequent states at \(\theta=0.558^\circ\) and \(\theta=1.054^\circ\). Insets: \(\SI{100}{nm}\times\SI{100}{nm}\) PFM images.
  \textbf{(E–H)} Theory–experiment comparison: Hexagon-averaged PFM maps (E,G) and Pseudomagnetic field textures computed from fitted lattice parameters (F,H) at \(\theta=0.100^\circ\) and \(\theta=0.558^\circ\).
  }
\end{figure*}

\textbf{Experimental setup.}
Figure~\ref{fig:Schematic} provides a schematic illustration of the rotor geometry and manipulation process in TBG system. First, a heterostructure of twisted bilayer graphene is assembled by mechanical exfoliation and dry transfer with a hBN substrate \cite{dean2010boron,edgecontacts2013}. Next, a lithographically defined metal rotor (roughly $5\si{\micro\meter}$ by $10$\si{\micro\meter}) is patterned on the top monolayer (Fig.~\ref{fig:Schematic}A), providing rigid coupling between the rotor and the active layer. Because adhesion to the rotor far exceeds interlayer friction and van der Waals forces between the two atomic layers, the rotor and top graphene layer translate and rotate together as a single body. Therefore applying a lateral force above a certain threshold moves the top graphene layer relative to the bottom layer. Before manipulation, standard contact-mode AFM cleaning is performed on the rotor surface \cite{goossens2012mechanical,lindvall2012cleaning}, which effectively removes most organic residues and reduces the friction for the rotation process. During the manipulation, the sample stage is heated, in order to further reduce interlayer friction.

Manipulation is executed with the AFM in contact mode using programmed, line-scan paths targeted to a chosen location on the rotor (Fig.~\ref{fig:Schematic}A-B). The rotor geometry allows for two distinct manipulation modes: a rotation mode, which applies torque to adjust the twist angle between layers, and a lateral translation mode, which induces in-plane translation without rotation. In the rotation mode, the tip scan direction is oriented to apply a longitudinal push perpendicular to the lever arm, setting both the direction and magnitude of the induced rotation. Typical vertical forces applied on the tip for initiating \textit{in situ} rotation in graphene rotor devices are \SIrange{600}{2000}{\nano\newton}. 
As illustrated in Fig.~\ref{fig:Schematic}C, placing the contact point far from the geometric center of the rotor applies a torque that controllably rotates the rotor and the attached top layer graphene. As we quantify below, this rotation changes the moiré wavelength nearly isotropically, while the rotor provides mechanical support that minimizes stretching or tearing. The relative angle change of the rotor is shown by overlaying “before/after" tapping-mode AFM images of the full frame (Fig.~\ref{fig:Schematic}C). 

Fig.~\ref{fig:Schematic}D exhibits the lateral translation mode, where near-pure translation is achieved by aligning the push direction with the contact point and the center of mass of the rotor. While AFM cleaning and heated sample stage reduce friction, there is still sliding resistance arising from interlayer friction and device topography, which prevents pure translation but remains sufficiently small to maintain predominantly translational motion. Combining these two modes, this technique allows precise and controllable changes to the relative position and angle between the two graphene layers, allowing for \textit{in situ} selection of the superlattice size.\\


\begin{figure*}
\includegraphics[width= 0.9\textwidth]{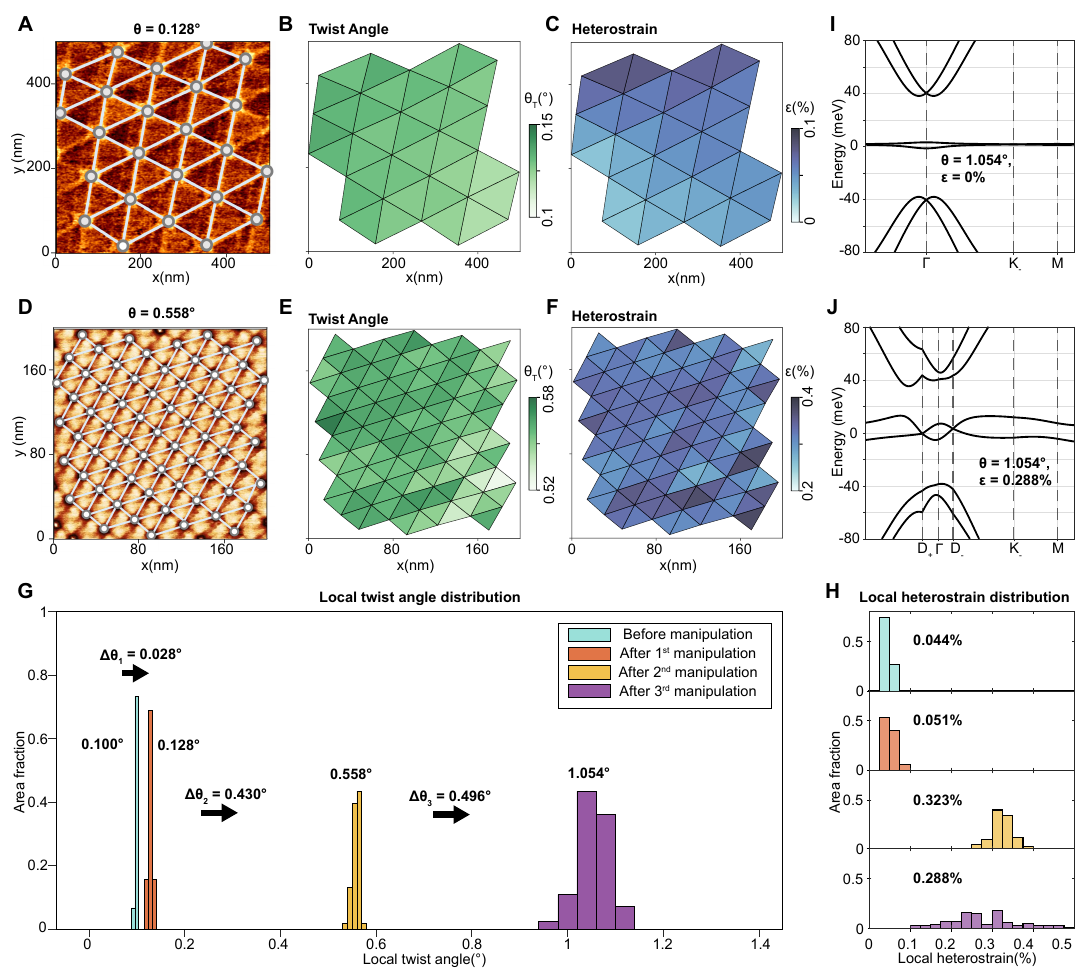}\caption{\label{fig:fig3} \textbf{Local disorder and geometry analysis of the graphene moiré superlattice} 
\textbf{(A-F)} Local twist angle and heterostrain maps extracted from PFM images of the moiré superlattice. Panel (A,D) are PFM images of the graphene moiré superlattice at $\theta=0.128\degree$ and $\theta=0.558\degree$, respectively. AA stacking sites (white dots) are used to define triangular units (grid lines), from which spatial maps of the local twist angle $\theta_T$ (B,E) and heterostrain $\epsilon$ (C,F) are extracted.
\textbf{(G-H)} Histograms of $\theta_T$ and $\epsilon$ across all four twist angle stages. The narrow spreads in both $\theta_T$ and $\epsilon$ confirm that successive manipulation preserve lattice quality with minimal added disorder.
\textbf{(I-J)} Calculated band structures near charge neutrality for $\theta=1.054\degree$, with $\epsilon=0\%$ and $\epsilon=0.288\%$ respectively. Heterostrain lifts degeneracies and modifies the bandwidth.}
\end{figure*}


\textbf{\textit{In situ} rotation of twisted bilayer graphene.}
We demonstrate \textit{in situ} dynamic rotation of a single monolayer graphene layer within a twisted bilayer graphene (TBG) heterostructure. Applying the rotation mode manipulation as above, we achieved a series of discrete changes in the superlattice periodicity, shown in Fig.~\ref{fig:fig2}A-D. The device was tuned to four distinct twist angles: $0.1^\circ$, $0.128^\circ$, $0.558^\circ$, and $1.054^\circ$, enabling microscopic studies from the domain-wall-network regime to the magic-angle regime within a single device. The smallest angle increment we demonstrate is $\Delta \theta = 0.028\degree$, achieved by moving the rotor \(\sim\)\SI{30}{\nano\meter} at the AFM push point (Fig.~\ref{fig:fig2}A–B).
 The resulting twist angles can be read out with high precision using piezoresponse force microscopy (details in Method). Crucially, PFM images from the same location on the heterostructure (Fig.~\ref{fig:fig2}A-D)
 demonstrate the rotor produces an essentially isotropic rotation of one monolayer relative to the other. This, in turn, induces an isotropic change in the superlattice period with each successive twist, up to a small error quantified below.

The superlattices produced by our \textit{in situ} twisting method are highly uniform, as demonstrated via direct microscopy. To resolve fine sub–unit-cell features, we average the PFM signal over many moiré unit cells to produce a hexagonal moiré unit cell (see SI for details). Figure~\ref{fig:fig2}(E) shows the averaged unit cell at $0.1\degree$, where a triangular network of domain walls emerges clearly. This observation is consistent with theoretical predictions that, at minimal twist angles, lattice relaxation drives a reconstructed state characterized by a triangular domain-wall network, giving rise to quasi-1D electronic phenomena \cite{SanJose2013PRBHelicalNetworks,Efimkin2018_HelicalNetwork_PRB,Tsim2020PRBPerfect1DChiral,Huang2018PRL}. In contrast, at $0.558\degree$ (Fig.~\ref{fig:fig2}(G)), the signal is largest near the center of the unit cell, while the domain walls appear weaker and more spatially broadened. This evolution can be attributed to the energetic preference of graphene for local AB (Bernal) stacking over AA stacking by ${\sim}$\SI{10}{meV\per\angstrom^2}~\cite{NamKoshino2017}. During lattice relaxation, AB domains expand, but must be separated by AA stacking domain walls (which may be thought of as solitons in the atomic offset) with a characteristic width of ${\sim}\SI{5}{nm}$~\cite{NamKoshino2017,PhysRevB.98.224102, KangVafekTBGLatticeRelaxation2025, DeBeule2025MarginalTwistRelaxation}. 
Because this width remains roughly constant, the domain walls occupy a smaller fraction of the unit cell at smaller twist angles, making the network appear sharper. However, the domain wall width becomes comparable to the moiré wavelength at larger twist angles, leading to broadened features and a less sharply defined network.

To quantitatively validate this interpretation, we compare the PFM images with lattice relaxation theory. We model the displacement of the atomic positions using a displacement vector field, $\v{r} \to \v{r}' = \v{r} + \v{u}(\v{r}')$, determined by standard elastic theory~\cite{carr2017twistronics, NamKoshino2017,KangVafekContinuumModels2023,KangVafekTBGLatticeRelaxation2025}. At a given twist angle (and heterostrain, discussed below), we solve the Euler-Lagrange equations to find $\v{u}(\v{r}')$, as described in the SI. To compare with the PFM maps, which are sensitive to the high-strain regions of the sample, we compute the strain-sensitive pseudomagnetic field $\mathcal{B} = \hat{z} \cdot \nabla \times \mathcal{A}$ with $\mathcal{A} \propto (u_{xx} - u_{yy}, -2 u_{xy})$ where $u_{\alpha \beta} = (\partial_\alpha u_{\beta} + \partial_\beta u_{\alpha})$ is the usual strain tensor.  In the marginally twisted device $\theta=0.100\degree$, the domain wall features are pronounced and closely match the calculated pseudomagnetic field texture (Fig.~\ref{fig:fig2}E–F). At the larger twist angle of  $0.558^\circ$, the calculated real space pattern similarly reproduces the dominant features observed experimentally (Fig.~\ref{fig:fig2}G–H), including the domain-wall texture and the chiral asymmetric high-amplitude features inside each domain. However, some discrepancies remain between theory and experiment, including the the finer features surrounding each AA site. These differences may arise from deviations between $\mathcal{B}$ and the true piezo response, suggesting the presence of additional electro-mechanical mechanism. These results demonstrate continuous, single-device access to the relaxation-driven structural crossover in TBG, with direct microscopic and quantitative verification of lattice reconstruction across the small-angle regime.\\

\textbf{Analysis of superlattice quality: mapping twist angle disorder and heterostrain.}
Fig.~\ref{fig:fig3} analyzes the superlattice quality across the \textit{in situ} twist-angle evolution from $0.100^\circ$ to $1.054^\circ$ using a series of PFM images (spanning many unit cells), which provide a direct measure of superlattice quality as the twist angle is tuned.
Crucially, this provides a sequence of spatially-resolved ``snapshots'' of the heterostrain profile and twist angle disorder in the evolving superlattice -- key microscopic structural quantities that determine the resulting electronic band structure. 
To extract these quantities, we identify each AA stacking center in the PFM data and reconstruct each moir\'e unit cell (Fig.~\ref{fig:fig3}A, D). From the geometry of each triangle, we extract (see SI) the local twist angles $\theta_T$, shown in Fig.~\ref{fig:fig3}B, E. Beyond the twist angle, we also extract the local uniaxial heterostrain $\epsilon$ in Fig.~\ref{fig:fig3}C, F. Heterostrain is significant because, after the twist angle, it has the largest effect on the bandstructure near the magic angle \cite{ZhenBi2019,Parker2021_PRL,Xie2019}. For instance, the bandwidth the narrow (``flat") bands near the magic angle increases five-fold under a heterostrain of only $\epsilon \approx 0.3\%$ (Fig.~\ref{fig:fig3}I, J).

Overall, the maps and histograms in Fig.~\ref{fig:fig3}B, C, E, F show high uniformity of both $\theta_T$ and $\epsilon$, confirming that successive manipulation preserves lattice quality with minimal added disorder. Across three successive rotations spanning the marginal-twist regime to the magic-angle regime, from $\theta \approx 0.1^\circ$ initially to $\theta \approx 1.05^\circ$ at the final stage, the twist-angle disorder remains suppressed to $\sigma_{\theta} < 0.033^\circ$ throughout. Furthermore, as summarized in Fig.~\ref{fig:fig3}H, the mean heterostrain is at most $\epsilon \approx 0.32\%$ at all angles, even in the marginally twisted regime where superlattice regularity is more difficult to achieve. Notably,  these disorder and strain value are approximately at the same level with devices fabricated using tear-and-stack method\cite{deJong2022,Uri2020_TwistAngleDisorder_Nature,Kerelsky2019_MaximizedInteractions_Nature,Kazmierczak2021_StrainTBG_NatMater}. 
These quantitative results highlight the unique advantages of the rotor platform, which combines post-fabrication twist angle tunability at the atomic layer level with direct access to real-space imaging for structural characterization and verification of superlattice and quality. Importantly, this validation indicates that our multi-step rotation protocol reliably tunes the twist angle while keeping heterostrain, twist angle disorder, and other forms of mesoscopic disorder low enough to enable superlattice physics at a chosen twist angle.

To assess reproducibility, we applied the same protocol to a second twisted bilayer graphene device, which exhibited similar rotation-induced changes in the moiré superlattice (see Fig.~\ref{fig:pfmGraphene1} in the Supplementary Information). The same device architecture and manipulation technique also operate in twisted hBN/hBN bilayers, where we demonstrate tunable ferroelectric domain structures. Representative devices and results are provided in the SI (see Figs.~\ref{fig:S8}-\ref{fig:S9}).\\

\begin{figure*}
\centering 
\includegraphics[width= 0.9\textwidth]{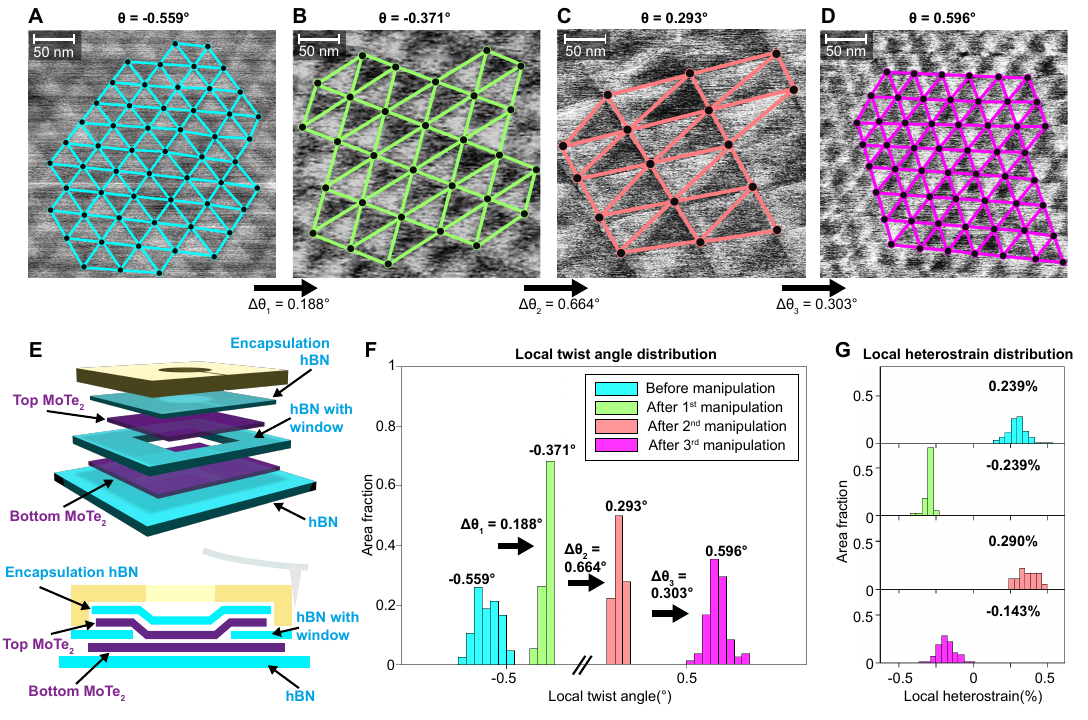}\caption{\label{fig:fig4}
\textbf{(A-D)} PFM images of the $\mathrm{MoTe_2}$ moiré superlattice before and after three successive \textit{in situ} AFM manipulations. Colored overlays highlight the domain walls of the moiré unit cell. 
\textbf{(E)} Three-dimensional schematics of the $\mathrm{MoTe_2}$ rotor. The encapsulation hBN protects the $\mathrm{MoTe_2}$ from air, while an etched window in the middle hBN allows the two $\mathrm{MoTe_2}$ layers to directly touch and form the moiré superlattice. 
\textbf{(F-G)} Histograms of the local twist angle and heterostrain across all four twist angle stages.
}
\end{figure*}

\textbf{\textit{In Situ} Twisting of \ce{MoTe2}.}
The same \textit{in situ} twisting method is easily adapted to manipulate single atomic layers in homobilayer systems beyond graphene, including air-sensitive TMDs. To demonstrate this, we realized a fully twistable, hBN-encapsulated $\mathrm{MoTe_2}$ homobilayer rotor device.
To protect the $\mathrm{MoTe_2}$ from oxidation while preserving the rotatable structure, we introduce hBN encapsulation layers with central openings, enabling both structural protection and direct imaging of the twisted region. The detailed fabrication procedures are provided in the SI.

Using the same AFM-based manipulation technique developed for graphene rotors, we achieve multiple \textit{in situ} twist-angle adjustments in a $\mathrm{MoTe_2}$ rotor. Fig.~\ref{fig:fig4}A–D show PFM images that track the evolution of the moiré superlattice through three rotations, during which the twist angle is step-wisely tuned from $-0.559^\circ$ to $0.596^\circ$. Fig.~\ref{fig:fig4}E shows the schematics of the twisted $\mathrm{MoTe_2}$ rotor. Applying the same analysis methods as in the graphene case, we extract the local twist angle and uniaxial heterostrain from each PFM image (spatial maps shown Fig.~\ref{fig:silocalmaps_mote2} in the SI). Histogram analysis reveals that the manipulation process introduces minimal additional twist-angle disorder (Fig.~\ref{fig:fig4}F), with the standard deviation slightly decreasing from $0.036^\circ$ in the initial state to $0.027^\circ$ after the third rotation. Similarly, heterostrain histograms across the four rotation stages show no significant increase in strain throughout the rotation process (Fig.~\ref{fig:fig4}G), from $0.239\%$ to $-0.143\%$, indicating that the system remains structurally stable throughout the manipulation process. Our heterostrain falls within, but toward the lower bound of, the range reported in prior STM characterizations of twisted $\mathrm{MoTe_2}$ devices \cite{osti_2899142, 10.1093/nsr/nwag014}.
To demonstrate reproducibility and the range of accessible angles, we implemented the same manipulation procedure on two additional devices, tuning each over a broad span of twist angles. Representative AFM and PFM datasets collected at each step are presented in the SI (Fig.~\ref{fig:SI_MoTe_rotor2}-~\ref{fig:SI_MoTe_rotor3}).\\

\textbf{Discussion and outlook.}
We have presented a robust route to dynamically tuning and imaging moir\'e superlattices by rotating one atomic layer relative to another \emph{in situ} using a commercial AFM. This approach achieves sub-degree angular control while minimizing induced heterostrain, preserving optical, transport, and scanning probe access, and enabling measurements at multiple twist angles within a single device. We demonstrate this workflow in graphene and extend it to air-sensitive TMDs, establishing a broadly applicable platform for the systematic study of angle-dependent phenomena.

Crucially, this capability grants access to the marginal twist regime --- a domain often challenging for fixed-angle assemblies. In minimally twisted graphene, lattice reconstruction and soliton networks strongly govern both local electronic structure and macroscopic response \cite{Alden2013PNAS,Huang2018PRL,Xu2019NatCommun}. However, heterostrain and spatial inhomogeneity often obscure flat-band physics at these small angles \cite{Kazmierczak2021NatMater,Mesple2021PRL}. By decoupling twist angle from initial assembly and enabling iterative angle verification via real-space moir\'e imaging, our approach provides a concrete experimental path toward low-strain, marginally twisted devices and the controlled exploration of reconstruction-driven transport networks \cite{Huang2018PRL,Yoo2019NatMater}.

In TMD moir\'e systems, continuous twist control directly targets a regime rich in strong correlations and topology. Twisted \ce{MoTe2} bilayers, for example, exhibit electrically tunable correlated magnetism \cite{Anderson2023Science} and signatures of integer and fractional quantum anomalous Hall states \cite{Park2023Nature, Cai2023Nature, Zeng2023Nature}. Given the critical role of polarization and stacking in shaping band topology \cite{Zhang2024NatCommun}, our platform is uniquely suited for mapping phase boundaries. It allows for the rigorous disentanglement of twist, strain, and electrostatic control within a single device, facilitating precise sweeps through the low-angle \ce{MoTe2} parameter space \cite{Anderson2023Science,Zeng2023Nature}.

Looking ahead, this method extends beyond the active moir\'e homobilayer. In graphene-based systems, the alignment between graphene and encapsulating hBN generates minibands and fractal quantum anomalous Hall physics, generate Chern bands resulting QAH physics \cite{Dean2013Nature,Hunt2013Science}, while in rhombohedral graphene, hBN alignment dictates ferroelectric and topological behaviors \cite{Lu2024Nature,Winterer2024NatPhys}. Rotating an encapsulating hBN layer \emph{in situ} offers a novel knob to tune moir\'e potentials and examine the evolution of ferroelectric textures under continuously adjustable orientation \cite{Niu2022NatCommun,Lu2024Nature}. By leveraging standard nanofabrication and widely available AFM instrumentation, this technique accelerates high-precision mapping of the moir\'e phase space with unprecedented control over twist angle and disorder.

\textbf{Materials and Methods.} \textit{Fabrication of graphene and hBN rotors:} Graphene and hexagonal boron nitride (hBN) flakes were mechanically exfoliated onto \(\mathrm{Si/SiO_2}\) substrates with a \SI{285}{\nano\meter} thermal oxide using the Scotch-tape method. Candidate flakes were identified by optical contrast.

Heterostructures were assembled by a dry-transfer process using a polycarbonate (PC) / polypropylene carbonate (PPC) film on a polydimethylsiloxane (PDMS) stamp. Flakes were cut using an AFM tip and then sequentially picked up and released onto target substrates. After release, polymer residue was removed by chloroform soak followed by light contact-mode AFM cleaning with a soft cantilever (spring constant \(\sim\!2\,\mathrm{N/m}\)) at \(\sim\!10\,\mathrm{nN}\) normal load. Edges of thin hBN were avoided to prevent tearing.

Rotor frames (\SI{10}{\nano\meter} Ti + \SI{150}{\nano\meter} Au) were patterned by electron-beam lithography (EBL) on \(\sim\!500\,\mathrm{nm}\) PMMA (950K A6) E-beam resist. Metals were deposited by e-beam evaporation at a base pressure \(\sim\!3\times10^{-7}\,\mathrm{Torr}\). The rotor geometry is patterned to span most of the top-layer perimeter, providing strong mechanical clamping and robust adhesion. Less than \SI{800}{nm} is left between the rotor frame and the flake edge to prevent metal deposition onto the bottom layer, which could otherwise pin and immobilize the top layer. Surrounding the bottom layer flake, metal is deposited along the edges to pin down the bottom layer flake. A final light contact-mode AFM clean removes lithography residuals. This architecture was used for graphene and for double-layer hBN rotor devices.

\textit{Fabrication of \texorpdfstring{MoTe\(_2\)}{MoTe2} rotors:} High-purity \(\mathrm{MoTe_2}\) (from HQ Graphene) was exfoliated onto \(\mathrm{Si/SiO_2}\) (\SI{285}{\nano\meter} oxide). Monolayers were identified by optical contrast. All \(\mathrm{MoTe_2}\) handling and assembly were performed in a nitrogen glovebox (O\(_2\), H\(_2\)O \(<\) \SI{0.1}{ppm}). Where needed, AFM tip cutting was used to define paired \(\mathrm{MoTe_2}\) regions for stacking.

A three-dimensional schematic illustration of the rotor device structure is presented in Figure \ref{fig:fig4}E, which is composed of three parts. The bottom part consists of a graphite flake overlaid with a boron nitride flake that serves as both the structural foundation and an electrostatic gate. The middle part is a TMD monolayer flake, encapsulated by a boron nitride layer that has an 5 × 5 $\mu m$ square-shaped opening created by reactive ion etching (RIE). The top part of the device comprises another TMD monolayer flake, which is encapsulated by two boron nitride flakes from the bottom and top. For this top part, there is an around 3 × 3 $\mu m$ opening on the bottom boron nitride flake, which enables effective contact between the central areas of the top and bottom TMD flakes through the aligned apertures. During the manipulation process, friction occurs between two hBN layers with etched windows. As reported in previous studies, the friction coefficient between angle-stacked hBN layers is predicted to be extremely low \cite{Hod2012PRB,Kabengele2021Nanoscale}, which facilitates the rotation of the entire top structure. Assembly of the device involves two sequential dry transfer steps. In the first step, the graphite back gate and the hBN base structure are transferred and deposited on a silicon substrate. Extensive AFM contact-mode cleaning is performed on this bottom part to ensure a clean and smooth surface for rotation. Following this, the rest of the structure is stacked, aligned, and deposited onto the base structure. After the heterostructure is made, a Ti/Au metal frame is deposited on top of the heterostructure using EBL. This metal frame is in contact with both hBN flakes which encapsulate the top TMD flake, which secures the relative positions of the top three layers and provides rigid structural support, allowing the hBN/TMD/hBN heterostructure to rotate as an integrated unit while minimizing strain induced by external force. 

\textit{PFM imaging:} Contact-resonance PFM (CR-PFM) was used for imaging the superlattice. the tip–sample contact resonance was identified and an AC drive (\(\sim\) \SI{1.5}{V}) was applied at resonance. Because the hBN/\(\mathrm{MoTe_2}\) moiré period is \(<\!1\,\mathrm{nm}\) (below CR-PFM resolution), the observed contrast originates from the twisted \(\mathrm{MoTe_2}\) superlattice. All imaging of \(\mathrm{MoTe_2}\) rotors are performed in ambient environment.

\textit{Method for \textit{in situ} rotation and translation:} Manipulation was performed with the AFM in contact mode using programmed, high–aspect-ratio (often single-line) scan paths targeted to a selected lever-arm location on the rotor. The scan direction was oriented to apply a longitudinal push approximately perpendicular to the lever arm, which sets both the sign and magnitude of the induced rotation. The push distance can be finely controlled by pausing the tip mid–line scan. For the manipulation, we used high spring constant $~45 N/m$ tapping-mode tips for higher lateral force. Typical normal loads for \textit{in situ} rotation in graphene rotor devices were \(\sim\!600\text{–}2000\,\mathrm{nN}\), with device and tip dependent variations. Loads up to \(3500\,\mathrm{nN}\) were used for \(\mathrm{MoTe_2}\) stacks under encapsulation. In each manipulation step, tapping-mode z height lines scans are used to verify movement of the rotor frame. Contact resonance PFM images were then collected to quantify the twist from the moiré wavelength and its local variations. The commanded scan path and push-point radius set the coarse \(\Delta\theta\). The PFM-derived periodicity provided the precise \(\Delta\theta\) used in analysis. The balance between rotation and translation is set by the push-point position: near-pure translation was achieved by aligning the contact point with the rotor’s center of mass (Fig.~\ref{fig:Schematic}D). To achieve the sequence of rotations featured in Fig.\ref{fig:fig2}A-D, the sample stage was heated up to $190\celsius$ to lower the interlayer friction between the graphene flakes, while longitudinal forces ranging from $\SI{1000}{nN}$ to $\SI{1500}{nN}$ were applied to the rotor frame. The smallest twist step we demonstrate is $\Delta \theta = 0.028\degree$, achieved with a \(\sim\)\SI{30}{\nano\meter} push at the rotor arm (Fig.~\ref{fig:fig2}A–B). The discrepancy between the projected rotation angle ($\Delta \alpha = 0.1^\circ$) and the twist angle change ($\Delta \theta = 0.028^\circ$) extracted from PFM images might be attributed to deformation of the edge of the frame at the contact point, which introduces an offset to the actual travel distance of the tip. AFM image of the rotor device used for these measurements is provided in Supplementary Fig. ~\ref{fig:pfmbeforeebeam}.

\bibliography{apssamp}

\section*{\label{sec:acknowledgement} Acknowledgments}

\noindent\textbf{Funding:}
We gratefully acknowledge funding support from the UC Office of the President, specifically the UC Laboratory Fees Research Program (award LFR-20-653926) and the AFOSR Young Investigator Program (award FA9550-20-1-0035). This work was performed, in part, at the San Diego Nanotechnology Infrastructure (SDNI) of UCSD, a member of the National Nanotechnology Coordinated Infrastructure, which is supported by the National Science Foundation (Grant ECCS-2025752).


\noindent\textbf{Author contributions:} Q.Z., L.L., D.E.P., and M.T.A. wrote the manuscript, with input from all authors. Q.Z., L.L., T.S., R.Z., L.W.C., and A.D. prepared the graphene rotor devices. Q.Z., L.L., S.P., and Z.Y. prepared the \ce{MoTe2} devices. Q.Z., L.L., and J.T. performed AFM and PFM of the rotor devices. Q.Z., C.W., and D.E.P. analyzed the data. K.W. and T.T. grew and provided the hBN crystals.

\noindent\textbf{Competing Interests:} The authors declare that they have no competing interests.

\noindent\textbf{Data and materials availability:} All data and data processing code are available upon request.

\onecolumngrid

\vspace{0.3cm}

\newpage
\begin{center}
\Large{\bf Supplementary Information for: Dynamic twisting and imaging of moiré crystals}
\end{center}

\supplementarysection

\subsection*{Geometric estimate of the rotation angle from AFM actuation}

Our primary angle metrology is obtained from PFM by extracting the moiré periodicity before and after each step. As an independent geometric cross–check, we estimate a projected rotation from the AFM actuation geometry using the frame displacement and the rotor lever–arm radius. Let \(R\) denote the lever–arm radius (distance from the instantaneous rotation center to the push point) and \(L\) the linear displacement of the push point between the “before” and “after” states. The two positions define a chord on a circle of radius \(R\), the enclosed central angle is
\begin{equation}
\Delta\theta_{\mathrm{geom}} \;=\; 2\,\arcsin\!\left(\frac{L}{2R}\right),
\label{eq:chord}
\end{equation}
which reduces to \(\Delta\theta_{\mathrm{geom}}\!\approx L/R\) for \(L\!\ll\! R\).

When the rotation is a few degrees, we coarsely register the change by overlaying the “before’’ and “after’’ tapping–mode AFM images of the full frame (Fig.~\ref{fig:1Dlinescan}A–C). For sub–degree rotations, we extract \(L\) from AFM height linecuts taken along the push direction at the contact point (Fig.~\ref{fig:1Dlinescan}D,E) by measuring the lateral shift between the two traces. The lever arm \(R\) is determined from the same images by locating the instantaneous rotation center and measuring the distance to the push point. In practice, the apparent center of rotation lies near the frame region opposite the push point and can shift with interfacial friction and local compliance.

\subsection*{Extraction of local twist angle and heterostrain from PFM images}

This subsection describes how the twist angle, heterostrain, and other parameters are extracted from PFM images. We first describe our geometrical conventions, then discuss the extraction procedure.

Let the lattice for the monolayer (either graphene or a TMD) be
\[
\mathbf{a}_1 = a\!\left( -\tfrac{1}{2},\, \tfrac{\sqrt{3}}{2} \right),
\qquad
\mathbf{a}_2 = a(-1,\,0),
\]
with corresponding reciprocal lattice vectors
\begin{equation}
\mathbf{G}_1 = \frac{2\pi}{a}\!\left( 0,\,\frac{2}{\sqrt{3}} \right),\quad
\mathbf{G}_2 = \frac{2\pi}{a}\!\left( -1,\,\frac{1}{\sqrt{3}} \right),\quad
\mathbf{G}_3 = -\mathbf{G}_1 - \mathbf{G}_2.
\end{equation}
These are related by $C_3$ symmetry: $C_3 \mathbf{G}_j = \mathbf{G}_{j+1 \pmod 3}$. The leading order distortion to the moir\'e unit cell is controlled by the heterostrain, which is described by a symmetric \(2\times 2\) tensor~\cite{ZhenBi2019}
\begin{equation}
\mathsf{S} \;=\;
\mathsf{R}_{-\phi}\!
\begin{pmatrix}
-\epsilon & 0\\[2pt]
0 & \nu\,\epsilon
\end{pmatrix}
\mathsf{R}_{\phi},
\qquad
\mathsf{R}_{\phi} =
\begin{pmatrix}
\cos\phi & -\sin\phi\\[2pt]
\sin\phi & \cos\phi
\end{pmatrix},
\end{equation}
where \(\epsilon\) is the uniaxial heterostrain magnitude, \(\phi\) the strain-axis angle, and \(\nu \approx 0.16\) the Poisson ratio of graphene. (Equivalently, \(\varepsilon_2=-\nu\,\varepsilon_1\) in principal axes.)

We consider a moir\'e pattern with twist angle \(\theta\). To first order in \(\theta\) and \(\epsilon\), the moiré lattice and reciprocal lattice are~\cite{ZhenBi2019}
\begin{equation}
\mathbf{L}_j = \mathsf{T}^{-T}\mathbf{a}_j,\qquad
\mathbf{g}_j = \mathsf{T}\,\mathbf{G}_j,\qquad
\mathsf{T} = \mathsf{S} + \theta\,\mathsf{J},
\quad
\mathsf{J}=\begin{pmatrix}0&1\\[2pt]-1&0\end{pmatrix},
\end{equation}
where \((-T)\) denotes inverse transpose and \(j=1,2,3\). Overall the moir\'e lattice is a function of the twist angle $\theta$, the heterostrain magnitude $\epsilon$ and heterostrain direction $\phi$. Finally, there is a global rotation between the lab frame and the coordinate system here, which is parametrized by a rotation angle $\alpha$ as $\v{L}_j^{\mathrm{lab}} = R_{\alpha} \v{L}_j$.

We now describe how $(\theta,\epsilon,\phi,\alpha)$ are extracted. From each PFM image we (i) identify AA-stacking sites and record their coordinates; (ii) choose two moiré vectors $(\mathbf{R}_1,\mathbf{R}_2)$ based on the coordinates to fix a local orientation; (iii) find similar vector pairs to identify moiré vectors for all triangle cells. Each set of two vectors can be connected into a triangular domain.

For each accepted triangle $(\mathbf{t}_1,\mathbf{t}_2,\mathbf{t}_3)$ we form measured edge vectors
\[
\mathbf{R}^{(\mathrm{meas})}_1=\mathbf{t}_2-\mathbf{t}_1,\quad
\mathbf{R}^{(\mathrm{meas})}_2=\mathbf{t}_3-\mathbf{t}_1,\quad
\mathbf{R}^{(\mathrm{meas})}_3=\mathbf{t}_3-\mathbf{t}_2.
\]
We then construct a triangle with parameters $p=(\alpha,\epsilon,\phi,\theta)$ (overall moiré lattice rotation, uniaxial heterostrain magnitude and axis, and twist angle) following the lattice geometry mentioned above. We then fit the parameters by minimizing the loss of the fitted triangle and the measured triangle:
\begin{equation}
    \mathcal{L}(\theta,\epsilon,\phi,\alpha) = \sum_{j=1}^3 \left|\left| \mathbf{R}^{(\mathrm{meas})}_j - \mathbf{L}_j(\theta,\epsilon,\phi,\alpha) \right|\right|^2.
\end{equation}

Repeating over all triangles produces spatial maps of $\theta_T$ and $\epsilon$.

\subsection*{Construction of the averaged hexagonal unit cell from PFM images}

We construct an averaged hexagonal unit cell from each PFM image. For consistency, we reorder the six vertices of each hexagon in a standard way (top corner first, then proceed counter-clockwise). Each selected hexagon is then unwrapped onto a common hexagon template that is generated using the fitted twist angle and strain profile calculated in previous step. We split the hexagon into six triangles that meet at the center and map each source triangle to the corresponding template triangle using a simple affine transform, combining them to produce one hexagon image on a fixed grid. Repeating this for all selected cells puts every unit cell into the same shape and orientation, after which we average them pixel-by-pixel to obtain the final mean hexagon. 

\subsection*{Lattice Relaxation Modeling}

This section describes the lattice relaxation modeling. We follow the elasticity theory approach of~\cite{NamKoshino2017} using the updated formulation in~\cite{KangVafekTBGLatticeRelaxation2025}.

We assume that the lattice relaxation is a continuous function of position in both layers $\ell = t,b$. Taking Eulerian coordinates, the distorted in-plane position $\v{X}_\ell$ is written in terms of the undisorted position $\v{r}$ as
\begin{equation}
	\v{X}_{\ell} = \v{r}+\v{u}_{\ell}(\v{X}_\ell).
\end{equation}
Out of plane corrugation is a subleading geometrical effect~\cite{KangVafekTBGLatticeRelaxation2025} (but enters the electronic structure with roughly equal magnitude to the in-plane distortions~\cite{TBornotTB,PhysRevB.108.094115}).

It is convenient to use symmetry and antisymmetric distortions $\v{u}^{+} = \left( \v{u}^t + \v{u}^b  \right)/2$ and $\v{u} = \v{u}^{-} = \v{u}^t - \v{u}^b$. The elastic theory for $\v{u}^{\pm}$ is described by a Lagrangian $L = T-U$ where $T$ is the standard kinetic energy and the potential energy $U = U_E + U_B$ is the sum of elastic energy $U_E$ and interlayer adhesion $U_B$. As we are interested in static configurations only ($T=0$), it is sufficient to simply minimize $U[\v{u}^{\pm}]$. In fact, $\v{u}^+ =0$ is minimized by a constant rotation and offset, so may be set to zero without loss of generality and we may focus on $\v{u} = \v{u}^{-}$ alone.

The elastic energy is the sum of the elastic and interlayer binding contributions~\cite{KangVafekTBGLatticeRelaxation2025}
\begin{equation}
	\label{eq:relaxation_functional}
U[\v{u}] =    \frac{1}{2}  \int d^2 \v{x}\; \left[ (\lambda+\mu) (u_{xx} + u_{yy})^2\right]
   + \mu \left[ (u_{xx} - u_{yy})^2 + 4 u_{xy}^2 \right] + V[\v{u}(\v{x})],
\end{equation}
where $\lambda$ and $\mu$ are Lam\'e factors and $u_{\alpha \beta} = (\partial_\alpha u_\beta + \partial_\beta u_{\alpha})/2$ is the strain tensor. The interlayer binding energy is a function of the offset $\v{\delta}$ between the lattices on the two layers (local rotation is neglected), and is thus periodic in the single layer lattice: $V[\v{d} +\v{G}]= V[\v{d}]$. For TBG, $V$ is minimized for AB stacking and maximized for AA stacking. It is described as a Fourier series
\begin{equation}
    V[\v{d}] = \sum_{\v{G}} V^{\v{G}} \cos(\v{G} \cdot \v{d})
\end{equation}
where the sum runs over the first three ``shells" of reciprocal lattice vectors. The binding energy is $C_6$ symmetric, even for TMDs~\cite{RamosAlonso2025}. We thus take
\begin{equation}
	V^{\pm C_3^j \v{G}_1} = c_1 \mu, \quad 
	V^{\pm C_3^j(\v{G}_1 - \v{G}_2)} = c_2 \mu, \quad
	V^{\pm C_3^j (2\v{G}_1)} = c_3 \mu,
\end{equation}
where $c_{1,2,3}$ are dimensionless constants. To leading order, these are unchanged by heterostrain.

\begin{table}[h]
    \centering
    \begin{tabular}{lcccccc}\toprule
    \textbf{Material}
    & $a$
    & $\lambda$
    & $\mu$
    & $c_1$
    & $c_2$ 
    & $c_3$    \\ \midrule
    \ce{TBG} & $2.46$ (\si{\angstrom}) & $3.25$ (\si{meV\per\angstrom\squared}) & $9.035$ (\si{meV\per\angstrom\squared})  & $8.6 \times 10^{-5}$ & $-7.9 \times 10^{-6}$ & $-2.0 \times 10^{-6}$\\
    \bottomrule
    \end{tabular}
    \caption{Parameters for lattice relaxation. Parameters for TBG follow~\cite{KangVafekTBGLatticeRelaxation2025,carr2017twistronics}. 
    }
    \label{tab:lattice_relaxation_parameters}
\end{table}

Within each unit cell, the offset vector between the two layers covers all relative offsets. We thus parametrize the lattice relaxation as
\begin{equation}
	\v{u}(\v{x}) = \mathsf{T}^T \v{x} + \v{v}(\v{x}),
	\quad 	\v{v}(\v{x}+\v{L}_j) =  \v{v}(\v{x}),
	\quad \v{v}(\v{x}) = - \v{v}(-\v{x}).
\end{equation}
which is represented by a Fourier series
\begin{equation}
	\v{v}(\v{x}) = \sum_{\v{g}} \v{v}^{\v{g}} \sin(\v{g}\cdot \v{x}),
\end{equation}
where the sum runs over all moir\'e reciprocal vectors and $\v{v}^{\v{g}} = -\v{v}^{-\v{g}}$ are real numbers. In practice we take up to 13 shells of reciprocal vectors (though far fewer often suffice).

To minimize Eq.~\eqref{eq:relaxation_functional}, we set $\frac{\delta U}{\delta \v{v}(\v{x})} = 0$, which yields the differential equation~\cite{KangVafekTBGLatticeRelaxation2025}
\begin{equation}
    \frac{1}{2} \begin{bmatrix}
	\mu (\partial_x^2 + \partial_y^2) + (\lambda + \mu) \partial_x^2 & (\lambda + \mu) \partial_x \partial_y\\
		(\lambda + \mu) \partial_x \partial_y & \mu (\partial_x^2 + \partial_y^2) + (\lambda + \mu) \partial_y^2 
	\end{bmatrix}
	\begin{bmatrix} v_x(\v{x}) \\ v_y(\v{x})\\ \end{bmatrix} 
	+ \sum_{\v{G}} V^{\v{G}} \sin(\v{g}_{\v{G}}\cdot \v{x}  + \v{G} \cdot \v{v}(\v{x}))
	 \begin{bmatrix} G_x \\ G_y  \end{bmatrix} = 0,
\end{equation}
where $\v{g}_{\v{G}} = \mathsf{T} \v{G}$. Fourier transforming, this becomes a non-linear equation for $\st{\v{v}^{\v{g}}}$:
\begin{equation}
	\label{eq:lattice_relaxation_nonlinear_equation}
    \frac{1}{2} \begin{bmatrix}
	\mu (g_x^2 + g_y^2) + (\lambda + \mu) g_x^2 & (\lambda + \mu) g_x g_y\\
		(\lambda + \mu) g_x g_y & \mu (g_x^2 + g_y^2) + (\lambda + \mu) g_y^2 
	\end{bmatrix}
	\begin{bmatrix} v_x^{\v{g}} \\ v_y^{\v{g}} \\ \end{bmatrix} 
	=  \sum_{\v{G}} V^{\v{G}} F[\v{v},\v{G}]^{\v{g}}
	\begin{bmatrix} G_x \\ G_y  \end{bmatrix} \quad \forall \v{g},
\end{equation}
where
\begin{equation}
	F[\v{v}, \v{G}](\v{x}) = \sin(\v{g}_{\v{G}} \cdot \v{x} + \v{G} \cdot \v{v}(\v{x})) = \sum_{\v{g}} F[\v{v}, \v{G}]^{\v{g}} \, \sin(\v{g} \cdot \v{x}).
\end{equation}
As $\v{g}_{\v{G}} \cdot \v{L}_j = 2\pi n$ for some integer $n$, it follows from Eq.~\eqref{eq:relaxation_functional} that the left-hand side is periodic on the moir\'e lattice. Furthermore, since both $\v{x}$ and $\v{v}(\v{x})$ are odd functions, we can again use a sine series:
\begin{equation}
	\label{eq:inverse_fourier_integral}
	F[\v{v}, \v{G}]^{\v{g}} = \frac{1}{A} \int_{A} d^2\v{x} \; \sin(\v{g}_{\v{G}} \cdot \v{x} + \v{G} \cdot \v{v}(\v{x})) \sin(\v{g} \cdot \v{x}),
\end{equation}
where $A$ is the (area of the) moir\'e unit cell.

\begin{figure}
    \centering
    \includegraphics[width=\linewidth]{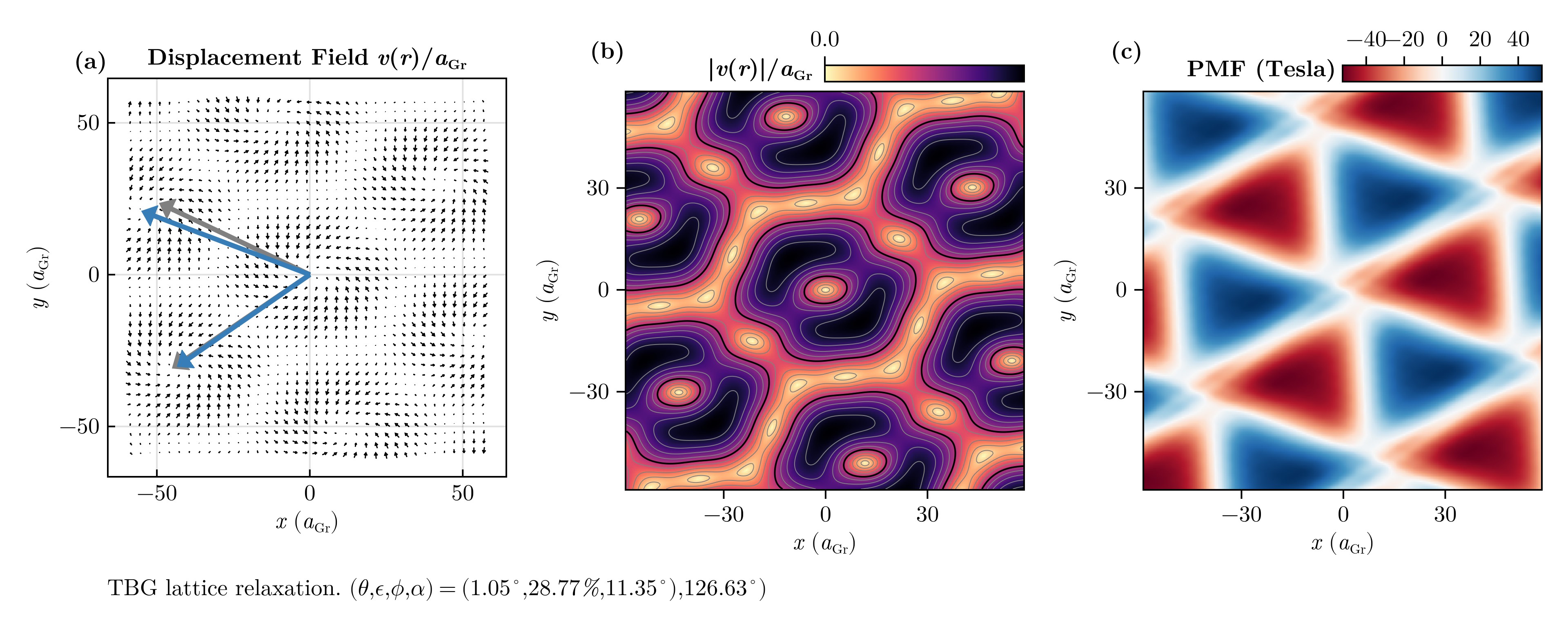}
    \caption{\textbf{Lattice relaxation in TBG.}
    (a) Displacement field $\v{v}(x)$. Blue arrows are the moir\'e lattice vectors in the presence of heterostrain and gray are their unstrained counterparts.
    (b) Magnitude of the displacement field $|\v{v}(\v{x})|$.
    (c) Resulting psuedomagnetic field.}
    \label{fig:lattice_relaxation_simulation}
\end{figure}

The non-linear equation Eq.~\eqref{eq:lattice_relaxation_nonlinear_equation} is easily solved by the standard iteration method: (1) start from $\v{v}(\v{x}) = 0$, (2) compute the right-hand side of Eq.~\eqref{eq:lattice_relaxation_nonlinear_equation} for each $\v{g}$ using the Fourier series Eq.~\eqref{eq:inverse_fourier_integral}, (3) compute the Fourier coefficients $\st{\v{v}^{\v{g}}}$ by solving the linear equation for each $\v{g}$, (4) repeat until convergence~\cite{NamKoshino2017,KangVafekContinuumModels2023}. To stabilize the convergence, it is often useful to take a finite ``learning rate" $v^{\v{g}}_{(n+1)} \leftarrow \eta v^{\v{g}}_{(n+1)} + (1-\eta) v^{\v{g}}_{(n)}$. For small angles of $0.1^\circ$, one should take $\eta<0.01$ or smaller.  In practice several hundred or thousand iterations are necessary, which takes about a second on a modern laptop. Convergence is exponential after an initial period, and could be accelerated using techniques such as DIIS or Anderson acceleration once one enters the regular regime. A representative relaxation structure is shown in Fig.~\ref{fig:lattice_relaxation_simulation}

Intriguingly, lattice relaxation spontaneously breaks chiral symmetry breaking and selects a particular handedness. This is because $F[-\v{v},\v{G}](\v{x}) \neq  -F[\v{v},\v{G}](\v{x})$ due to the offset from wrapping all possible offsets. For instance, one favors $v_y^{\v{g}_1} > 0$ over $v_y^{\v{g}_1} < 0$ in TBG near the magic angle.

The relationship between piezoelectric response and lattice relaxation has been studied by a number of works, but is still an active area of work in TBG and other centrosymmetric moir\'e materials~\cite{McGilly2020,Li2021,Zhang2024}. We opt to study a proxy for the PFM response that is expected to show features in similar locations in the unit cell: the psuedomagnetic field. This is defined in terms of the strain tensor as
\begin{equation}
 \mathcal{B}^{\mathrm{PMF}} = \nabla \times \mathcal{A} \text{ where }   \mathcal{A}(\v{r}) = \frac{3}{4} \frac{\beta \gamma_0}{e v_F} (u_{xx} - u_{yy}, - 2 u_{xy}).
\end{equation}

\subsection*{\textit{In Situ} twisting of double layer \ce{hBN}}

To demonstrate broad applicability to other classes of 2D materials, we extend this technique to double-layer hBN, which opens the door to \textit{in situ} control of the ferroelectric domain structure by tuning the twist angle \cite{zheng2020unconventional,woods2021charge,yasuda2021stacking}.

The fabrication procedure and the twisted heterostructure geometry are conceptually similar, as illustrated schematically in Fig. 1A. 
Fig. \ref{fig:S8}(A-B) show tapping-mode AFM images of the rotor configurations before and after a rotation, respectively. 

In double-layer hBN with a shallow twist angle near zero degrees, lattice reconstruction leads to the formation of a triangular network of AB and BA stacking domains with alternating electric polarizations \cite{zheng2020unconventional,woods2021charge,yasuda2021stacking}.
By tuning the twist angle between two individual monolayers of hBN, one can dynamically control the moiré ferroelectric domain structure, as revealed in the PFM images displayed in Fig. \ref{fig:S8}(C-D). 
To achieve the \textit{in situ} manipulation illustrated in Figure 4, a normal force of 2000 nN was applied to the sample surface while the tip traveled in the longitudinal direction. Upon contact with the sidewall of the rotor frame, the tip experienced torsional bending, generating a lateral force on the rotor frame. Meanwhile, the sample stage was heated to  190 \degree C to reduce interlayer friction. Using the same analysis method applied to the graphene moiré, we found the twist angle of the hBN superlattice was tuned from $0.1^\circ$ to $0.03^\circ$. In order to show the reproducibility of these results, additional data from a separate double-layer hBN rotor is provided in Fig. \ref{fig:S9}. 

\begin{figure}[htbp]
\includegraphics[width= 0.8\textwidth]{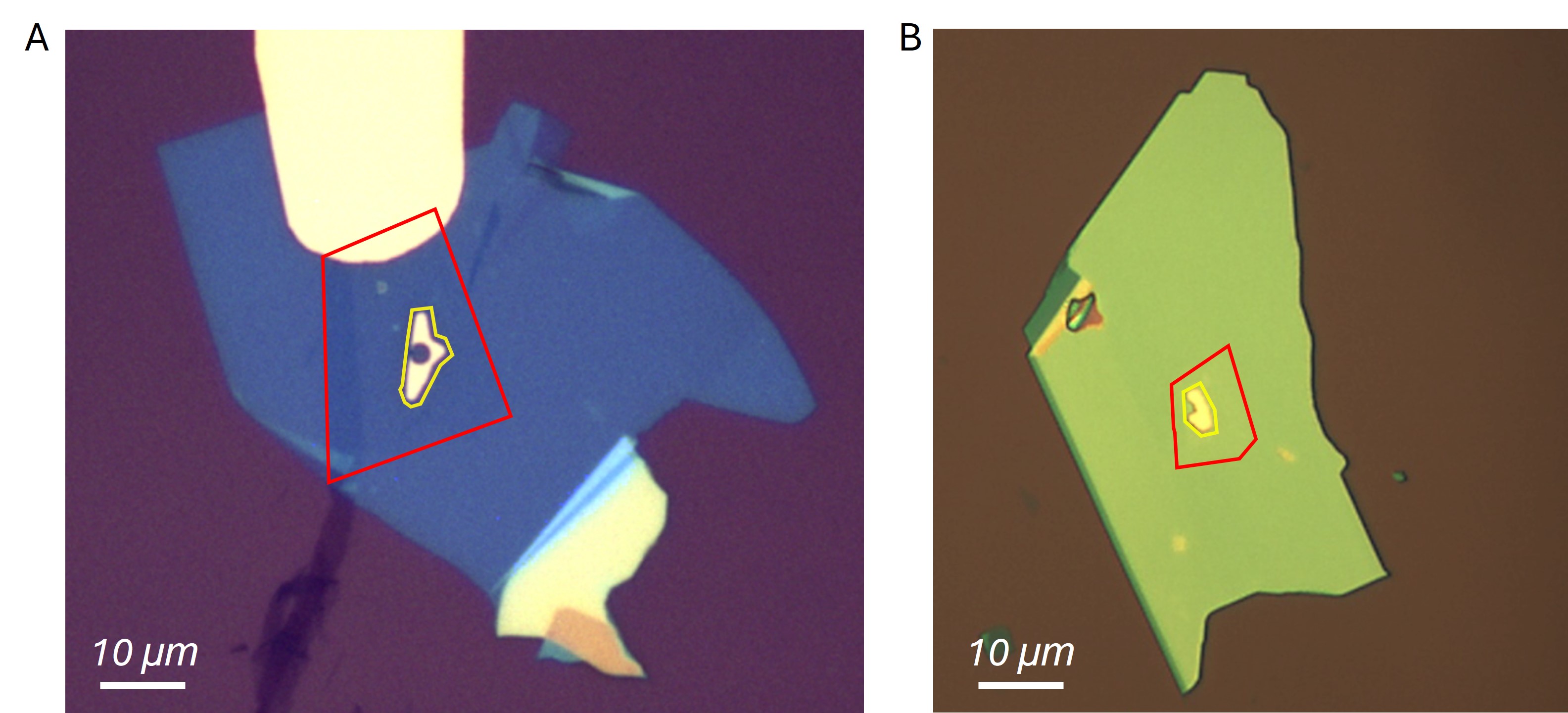}
\caption{\label{fig:opticalimages} \textbf{Optical images of rotor devices.} \textbf{(A)} Optical microscope images of the twisted bilayer graphene rotor deice 2 (featured in Fig.\ref{fig:pfmGraphene1}), contours of the two graphene layers are highlighted. \textbf{(B)} The double-layer hBN rotor device 2 (featured in Fig.\ref{fig:S9}), contours of the two hBN layers are highlighted. 
}
\end{figure}

\begin{figure}[htbp]
\includegraphics[width=\textwidth]{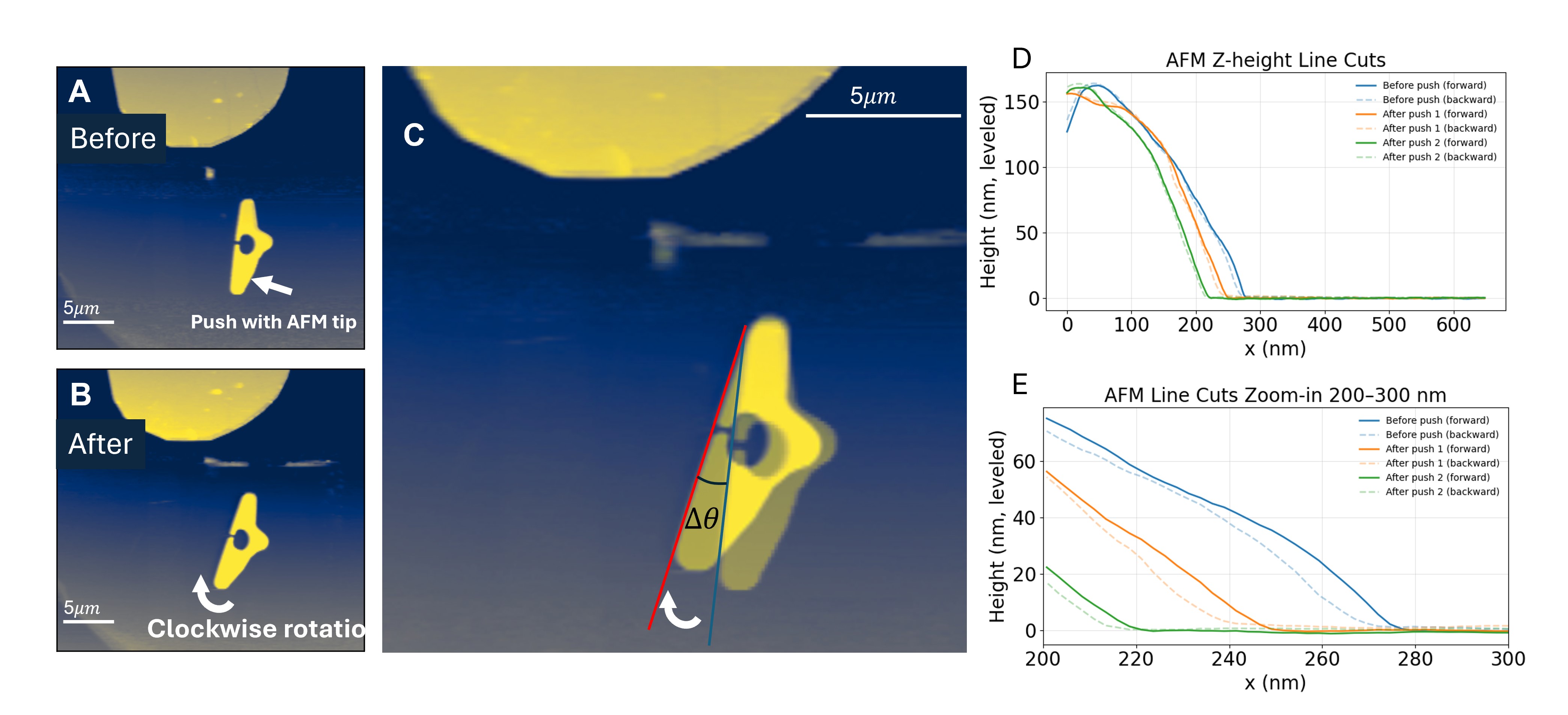}
\caption{\label{fig:1Dlinescan}
\textbf{Geometric approaches for estimating twist angle changes.}
\textbf{(A–C)} Tapping-mode AFM images of twisted bilayer graphene rotor acquired before (A) and after (B) a rotation. By overlaying the “before’’ and “after" images (C), the relative change angle is illustrated.
\textbf{(D,E)} For sub-degree rotations, AFM height linecuts taken before and after manipulation can be used to quantify the rotor displacement and estimate the angle change. In this example, comparison of the three linecuts indicates two consecutive \(\sim\!30~\mathrm{nm}\) shift along the push direction.
}
\end{figure}

\begin{figure*}
\includegraphics[width= 0.9\textwidth]{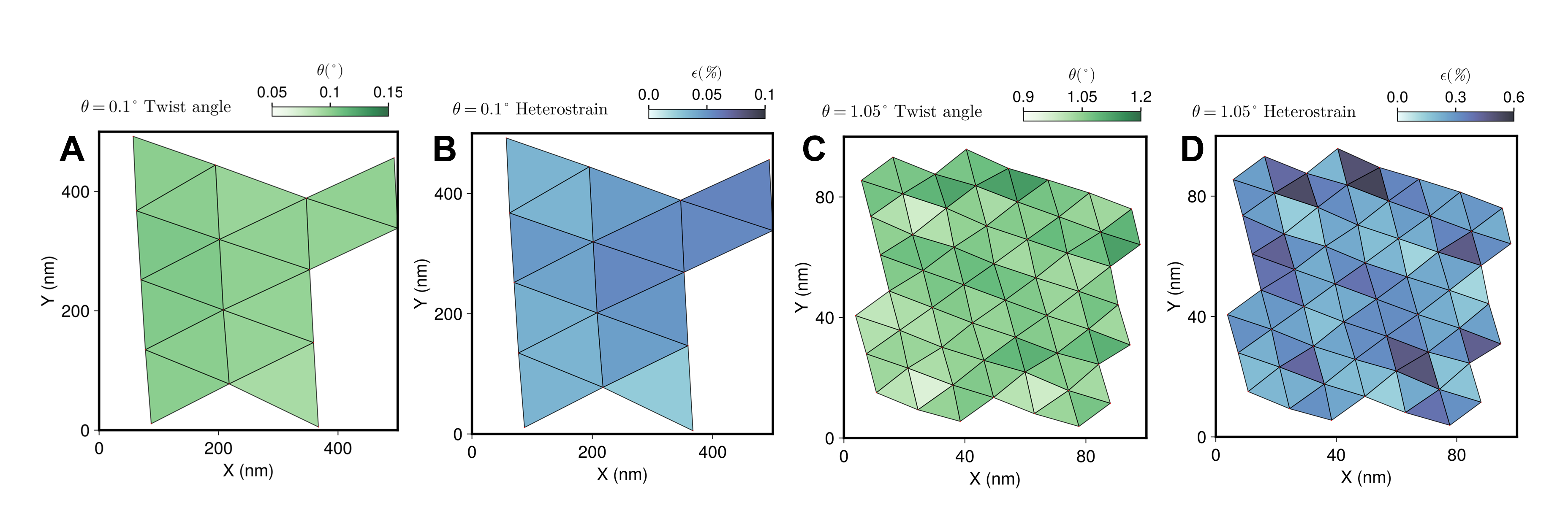}
\caption{\label{fig:silocalmaps_graphenerotor} \textbf{Local twist angle and heterostrain maps extracted from PFM images of the moiré superlattice for \(\theta=0.100^\circ\) (A-B) and \(\theta=1.05^\circ\) (C-D) in graphene rotor device 1, the device featured in Fig.\ref{fig:fig2}.}  
}
\end{figure*}

\begin{figure}[htbp]
\includegraphics[width= 0.7\textwidth]{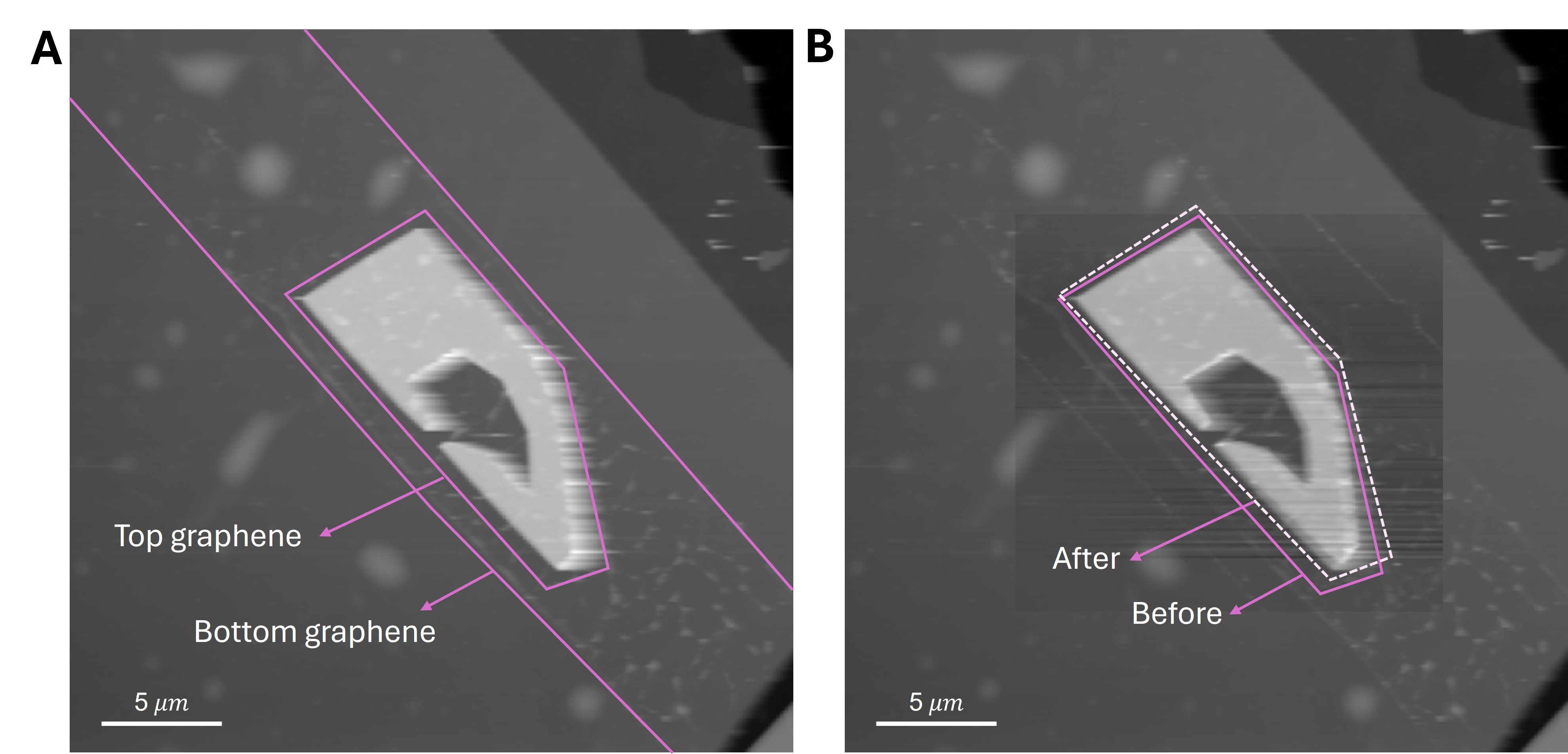}
\caption{\label{fig:pfmbeforeebeam} \textbf{AFM images of the rotor device (graphene rotor 1) featured in Fig. \ref{fig:fig2}.} (\textbf{A}) An AFM image of graphene rotor 1. (\textbf{B}) AFM images of the rotor prior to rotation and after the final rotation. The solid line and dashed line indicates the top graphene contour before and after the rotation, respectively.}
\end{figure}

\begin{figure}[htbp]
\includegraphics[width= 0.7\textwidth]{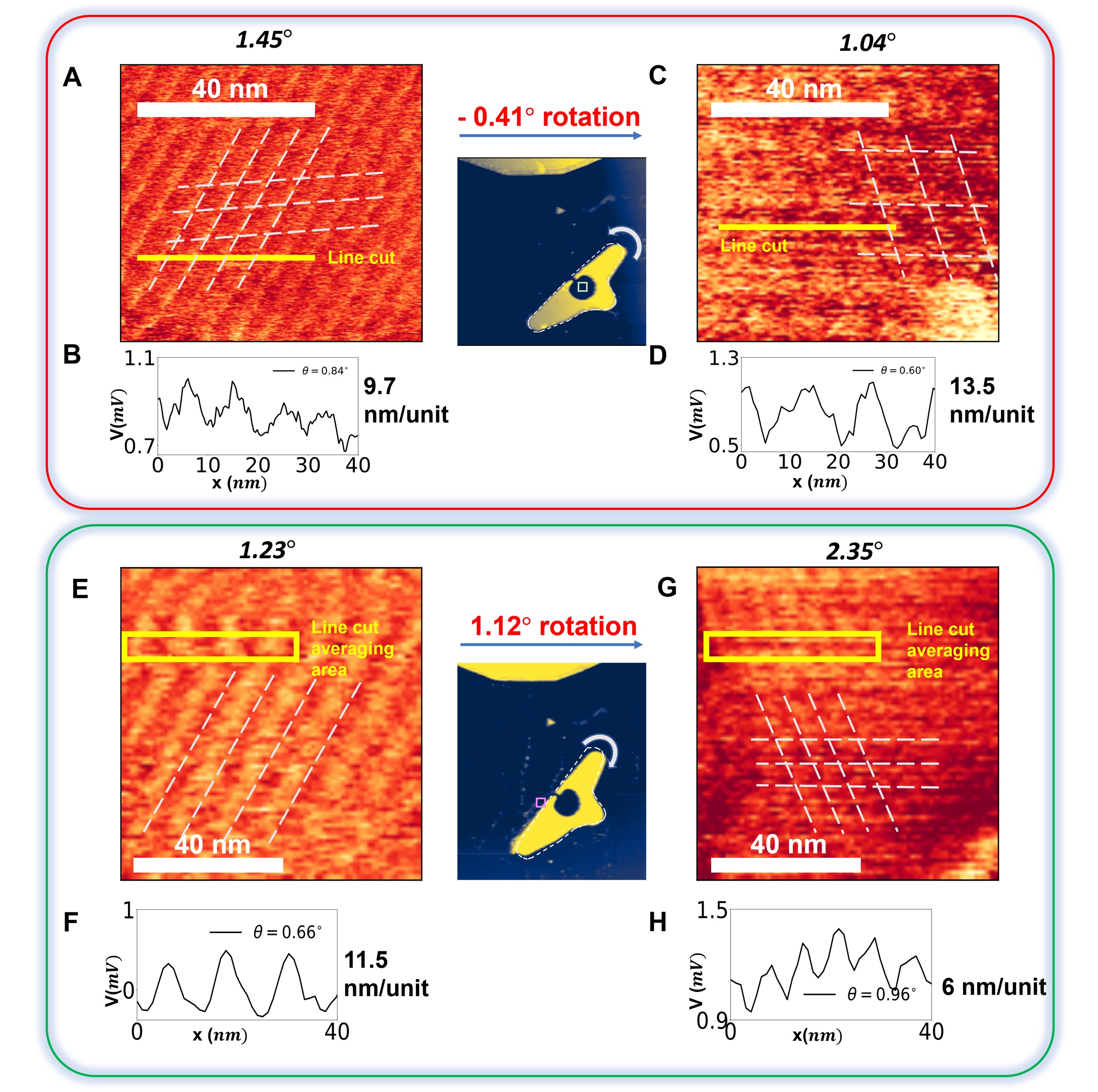}
\caption{\label{fig:pfmGraphene1} \textbf{Results from graphene rotor device 2.}
\textbf{(A, C)} PFM images of the twisted bilayer graphene rotor device before (A) and after (C) rotation, revealing the change of moiré superlattice period. The twist angle change is $\Delta \theta = -0.41\degree$. 
\textbf{(B, D)} 1D line cuts extracted from the PFM images in panels (A) and (C), respectively, showing the change in periodicity of the superlattice induced by the rotation.
\textbf{(E)} AFM image of the rotor device, illustrating the tip-induced manipulation responsible for the rotation of the twist angle.
\textbf{(E, G)} PFM images of a twisted bilayer graphene rotor device before (E) and after (G) a separate rotation process, showing the change in the moiré period that is caused by a twist angle change of $\Delta \theta = 1.12\degree$. 
\textbf{(F, H)} 1D line cuts extracted from the PFM images in panels (E) and (G), respectively, showing the change in periodicity of the superlattice induced by the rotation.
}
\end{figure}

\begin{figure}[htbp]
\includegraphics[width= 0.8\textwidth]{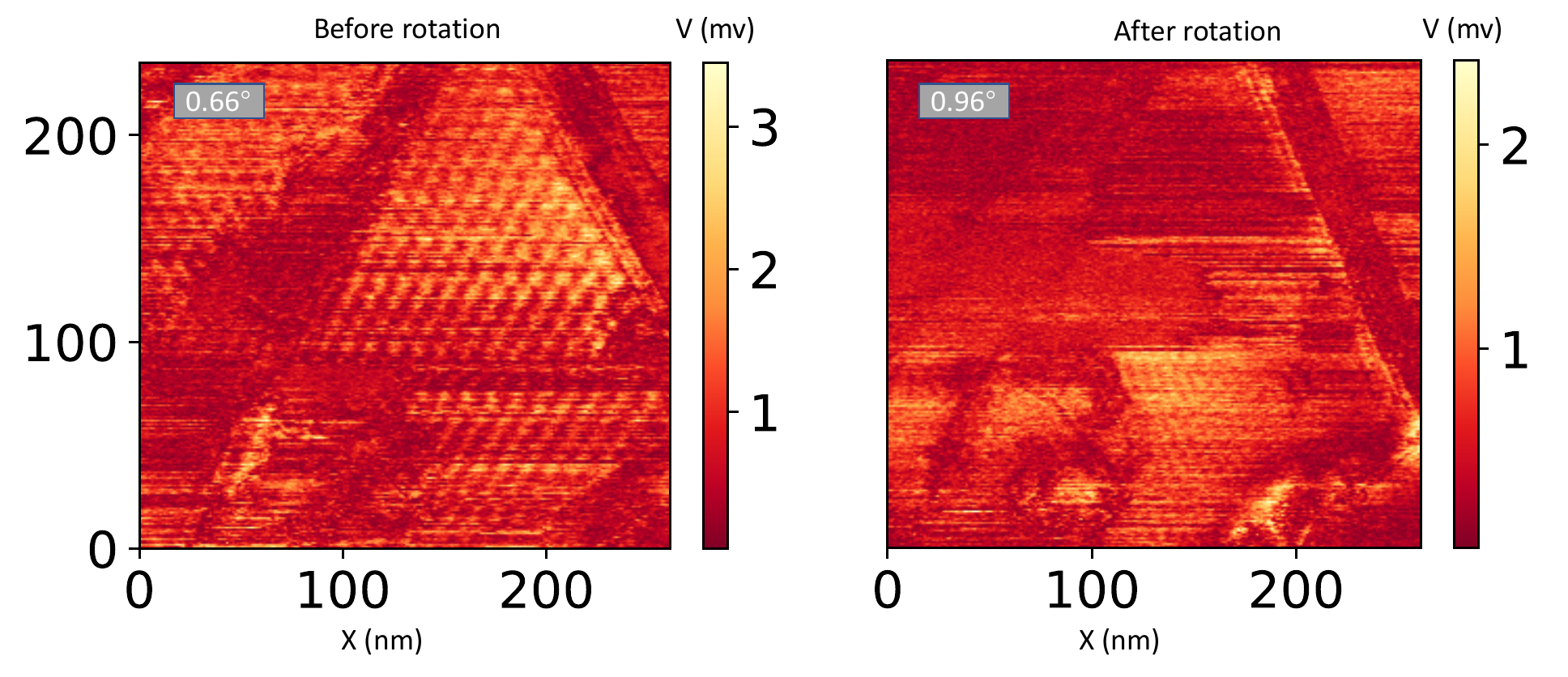}
\caption{\label{fig:fullpfm} 
\textbf{Partial delamination of unsupported graphene without a rotor frame.} 
The metal rotor frame mechanically supports the stack during rotation, clamping the top flake to the underlying heterostructure throughout the \textit{in situ} manipulation. Its importance is evident in the PFM images, which show partial delamination in the unsupported graphene region outside the frame (PFM acquired outside the rotor frame of graphene rotor 2; see Fig.~\ref{fig:pfmGraphene1}).
}
\end{figure}

\begin{figure*}
\includegraphics[width= 0.7\textwidth]{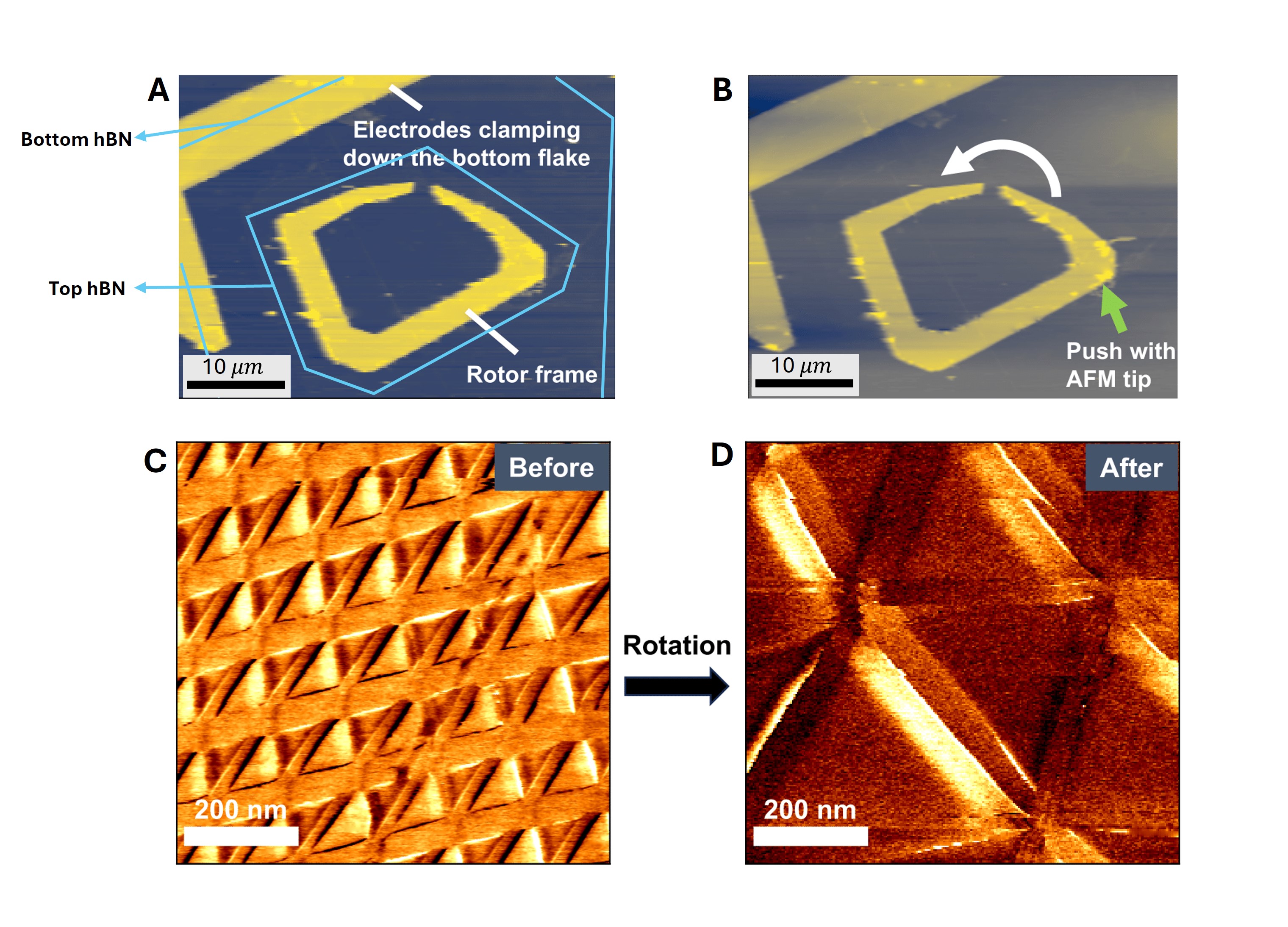}
\caption{\label{fig:S8} \textbf{Tuning the ferroelectric domain structure of twisted double layer hBN device 1.}
\textbf{(A-B)} AFM images of a twisted bilayer hBN rotor device before (A) and after (B) a rotation driven by a scanning probe tip. The green arrow indicates the push location on the rotor frame, while the white arrow marks the direction of the induced rotation. 
\textbf{(C-D)} PFM images showing the change of the ferroelectric domain structure in the twisted bilayer hBN induced by the \textit{{in situ}} rotation process. The twist angles of the heterostructure extracted from the moiré periodicity are 0.1\degree and 0.03\degree before and after rotation, respectively. The imaging and mechanical manipulation steps were performed within the same microscope.
}
\end{figure*}

\begin{figure*}
\includegraphics[width= 0.9\textwidth]{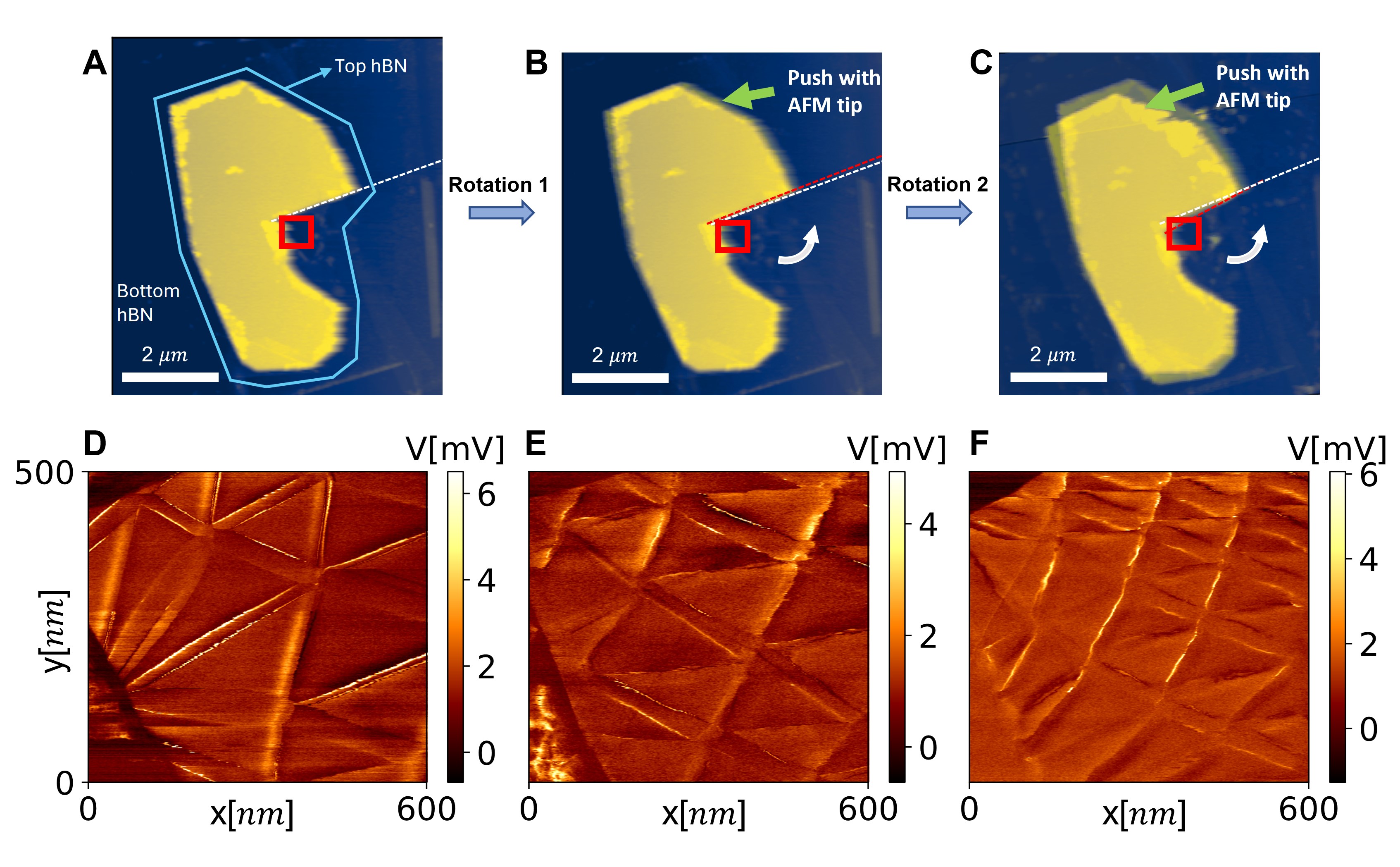}
\caption{\label{fig:S9} \textbf{Data from boron nitride rotor device 2: Tuning the ferroelectric domain structure of twisted double-layer hBN.}
\textbf{(A-C)} A sequence of AFM images of a second hBN rotor device before and after two successive rotations driven by the scanning probe tip. The white and red dashed lines mark the initial and final orientations of rotor, respectively.
The green arrows in (B) and (C) indicate the push point location and direction of the applied in-plane force, while the red squares mark the locations of the PFM imaging windows.
\textbf{(D-F)} Sequence of PFM images showing the evolution of the moiré domain structure in the twisted double-layer hBN induced by the mechanical rotation process. The imaging and \textit{in situ} manipulation steps were performed within the same microscope.
}
\end{figure*}

\begin{figure}[htbp]
  \centering
  \includegraphics[width=0.8\textwidth]{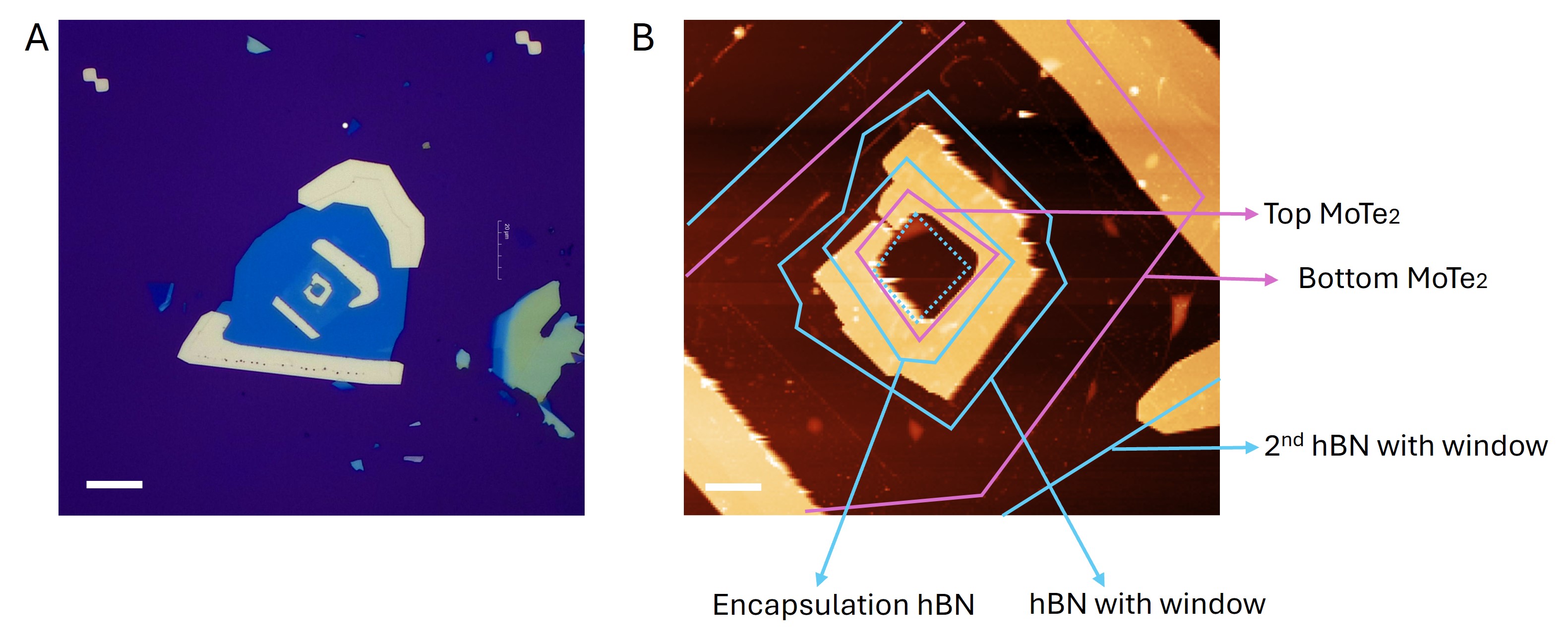}
  \caption{\label{fig:Mote2_rotor_optical}
  \textbf{Optical and AFM images of \(\mathrm{MoTe_2}\) rotor device 1.}
  \textbf{(A)} Optical micrograph of the \(\mathrm{MoTe_2}\) rotor device featured in Fig.~\ref{fig:fig4}. Scale bar: \(20\,\mu\mathrm{m}\).
  \textbf{(B)} AFM images of the device. Scale bar: \(5\,\mu\mathrm{m}\).
  }
\end{figure}

\begin{figure*}
\includegraphics[width= 0.9\textwidth]{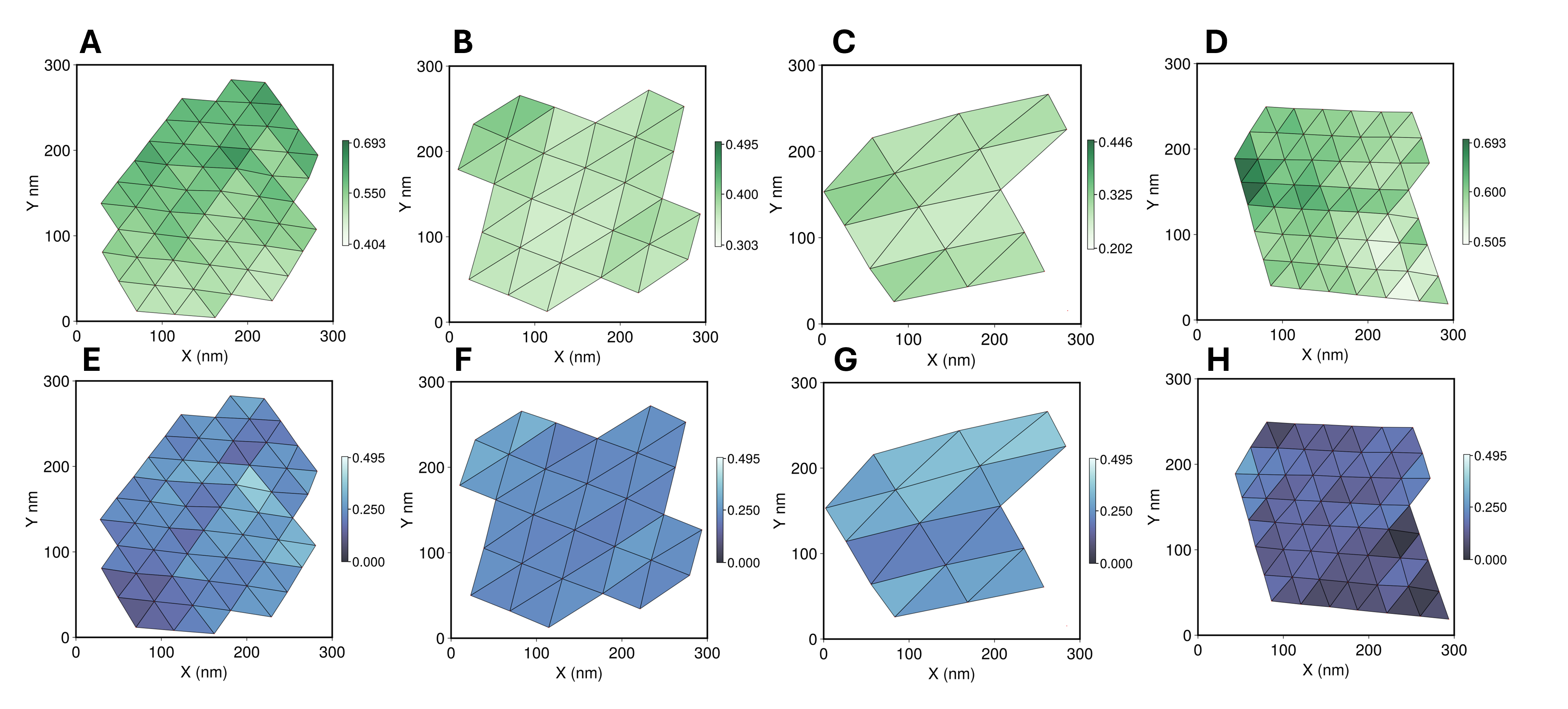}
\caption{\label{fig:silocalmaps_mote2} \textbf{Local twist angle and heterostrain maps extracted from PFM images of the moiré superlattice for the \(\mathrm{MoTe_2}\) rotor device 1 featured in Fig.~\ref{fig:fig4} } 
\textbf{(A–D)} Local twist angle maps for the \(\mathrm{MoTe_2}\) rotor device (unit:$^\circ$).
\textbf{(E–H)} Local heterostrain maps for the \(\mathrm{MoTe_2}\) rotor device (unit:\%).
}
\end{figure*}

\begin{figure*}
\includegraphics[width= 0.8\textwidth]{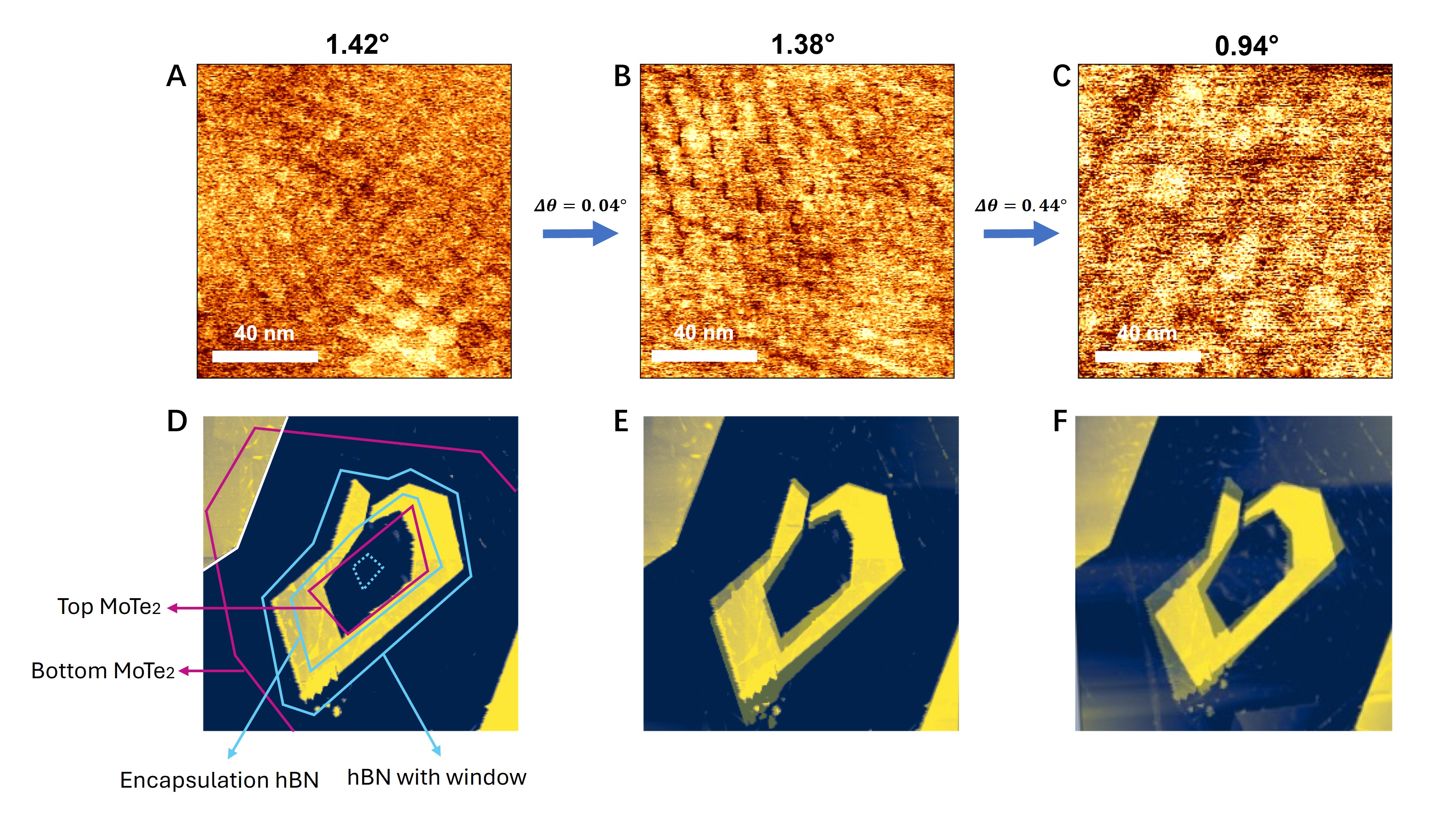}
\caption{\label{fig:SI_MoTe_rotor2}\textbf{Dynamic rotation and imaging of twisted bilayer MoTe\textsubscript{2} rotor 2.} 
\textbf{(A–C)} Piezoresponse force microscopy (PFM) images of the \(\mathrm{MoTe_2}\) moiré superlattice: (A) before rotation, (B) after the first rotation, (C) after the second rotation (twist angles are extracted from linecuts).
\textbf{(D–F)}  Tapping-mode AFM images of the rotor device: (D) before rotation, (E) after the first rotation, (F) after the second rotation, the outline from (D) is overlaid on (E,F) to highlight the positional changes of the rotor during manipulation.
}
\end{figure*}

\begin{figure*}
\includegraphics[width= 1\textwidth]{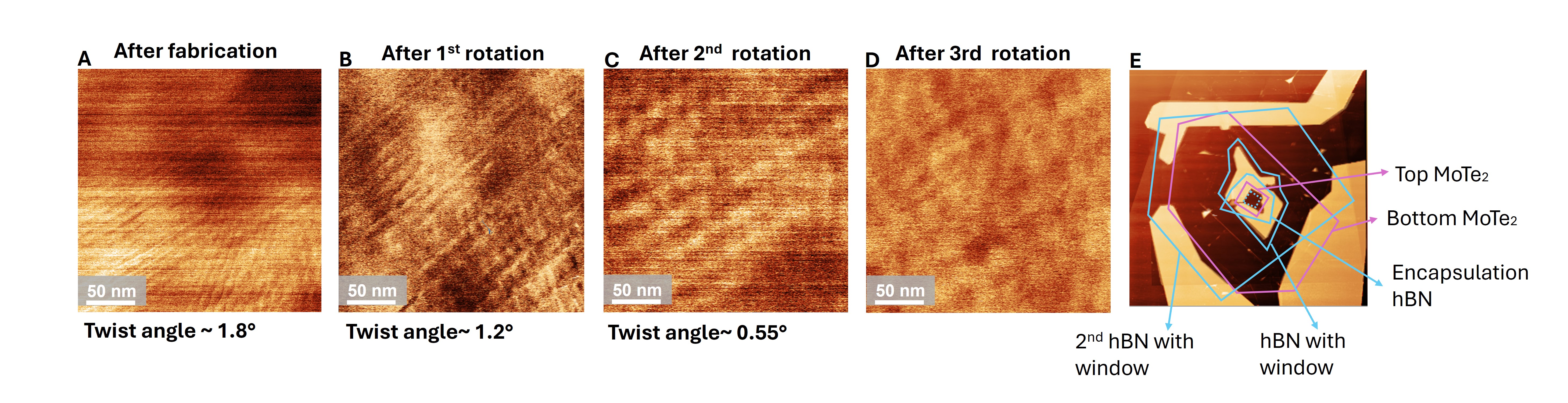}
\caption{\label{fig:SI_MoTe_rotor3} \textbf{Dynamic rotation and imaging of twisted bilayer MoTe\textsubscript{2} rotor 3.} 
\textbf{(A–D)} Piezoresponse force microscopy (PFM) images of the \(\mathrm{MoTe_2}\) moiré superlattice: (A) before rotation, (B) after the first rotation, (C) after the second rotation (D) after the third rotation. Although image quality is not high enough to support fine analysis, we could estimate the twist angle from linecuts of the PFM amplitude. The superlattice is tuned from around 1.8\degree to 0.55\degree after three consecutive rotations. \textbf{(E)} AFM images of the rotor device.}
\end{figure*}


\vfill

\end{document}